\documentclass[reprint,twocolumn,superscriptaddress, amssymb,aps,prl]{revtex4-1}
\usepackage{graphicx}
\usepackage{xcolor} 
\usepackage{amsmath}
\usepackage{hyperref}

\newcommand{\bra}{\left \langle}
\newcommand{\ket}{\right \rangle}

\begin{document}
\title{A Blue-Detuned Magneto-Optical Trap of CaF Molecules}

\author{Samuel J. Li}
\thanks{These authors contributed equally to this work.}
\affiliation{Department of Physics, Princeton University, Princeton, New Jersey 08544 USA}
\author{Connor M. Holland}
\thanks{These authors contributed equally to this work.}
\affiliation{Department of Physics, Princeton University, Princeton, New Jersey 08544 USA}
\author{Yukai Lu}
\affiliation{Department of Physics, Princeton University, Princeton, New Jersey 08544 USA}
\affiliation{Department of Electrical and Computer Engineering, Princeton University, Princeton, New Jersey 08544 USA}

\author{Lawrence W. Cheuk}
\email{lcheuk@princeton.edu}
\affiliation{Department of Physics, Princeton University, Princeton, New Jersey 08544 USA}

\date{\today}

\begin{abstract}
A key method to produce trapped and laser-cooled molecules is the magneto-optical trap (MOT), which is conventionally created using light red-detuned from an optical transition. In this work, we report a MOT for CaF molecules created using blue-detuned light. The blue-detuned MOT (BDM) achieves temperatures well below the Doppler limit, and provides the highest densities and phase-space densities reported to date in CaF MOTs. We observe short BDM lifetimes at high magnetic field gradients, preventing magnetic compression as a means to increase densities. By directly measuring the BDM restoring force, we find that the short lifetimes are explained by low effective trap depths. Notably, we find sub-mK depths at typical magnetic gradients, in contrast to $\sim 50\,\text{mK}$ depths in red molecular MOTs and $\sim0.5\,\text{K}$ depths in red atomic MOTs.  

\end{abstract}

\maketitle

Laser-cooled molecules are a promising platform for quantum simulation, quantum information processing, studies of ultracold molecular collisions, and precision probes of physics beyond the Standard Model~\cite{demille2002quantum,Carr2009review,Bohn2017molreview,blackmore2018reviewMoleculeQuantum}. These applications all benefit from trapped molecular samples that are cold and dense. A workhorse technology for producing trapped laser-cooled atoms and molecules is the magneto-optical trap (MOT), where a magnetic gradient in combination with near-resonant light provides both cooling and spatial confinement. Starting with the first molecular MOTs of SrF~\cite{barry2014SrFMOT,McCarron2015improvedMOT}, MOTs have been demonstrated in other diatomic and polyatomic molecules~\cite{Truppe2017Mot, Anderegg2017MOT, Collopy2018YOMOT, Vilas2022MOTCaOH}. These developments have enabled trapping of molecules in conservative optical and magnetic traps~\cite{Williams2018MagTrap, Anderegg2018ODT,Langin2021SrFODT,Wu2021YOODT,Anderegg2023trapCaOH}, the production of arrays of single molecules ~\cite{Anderegg2019Tweezer,Holland2023bichromatic}, explorations of molecular collisions in the ultracold regime~\cite{Cheuk2020collisions, Anderegg2021shield,Jorapur2023blueMOT}, and observations of coherent electric dipolar interactions and entanglement of molecules~\cite{Holland2022Tweezer,Bao2022Tweezer}.

Conventionally, molecular MOTs are created using light red-detuned from an optical transition. They primarily rely on Doppler cooling and achieve temperatures near the Doppler limit $T_D= \hbar \Gamma/(2k_B)$, where $\Gamma$ is the excited state linewidth. To achieve the optical cycling needed for laser-cooling, the cooling light addresses a rotational-lowering transition~\cite{stuhl2008Cycling} (e.g.  $N=1\rightarrow N'=0$ transition, where $N$ denotes the rotational quantum number). In these so-called type-II systems, the number of excited states is less than or equal to the number of ground states. As pointed out in \cite{devlin2016coolingSim}, with red-detuned light, type-II systems experience Doppler cooling at high velocities but sub-Doppler heating at low velocities. With blue-detuned light, Doppler heating and sub-Doppler cooling occur instead~\cite{devlin2016coolingSim,Truppe2017Mot,Cheuk2018Lambda,Ding2020YOSD,Langin2021SrFODT}. It was subsequently realized that blue-detuned MOTs (BDMs), which offer spatial confinement in addition to sub-Doppler cooling, are possible~\cite{tarbutt2018blueMOT, Langin2023simulation,Xu2022blueMOTsim}. The first BDM was realized with Rb atoms~\cite{tarbutt2018blueMOT}; recently, molecular BDMs of YO~\cite{Burau2023BlueMOT} and SrF~\cite{Jorapur2023blueMOT} have been created.

\begin{figure}[t]
	{\includegraphics[width=\columnwidth]{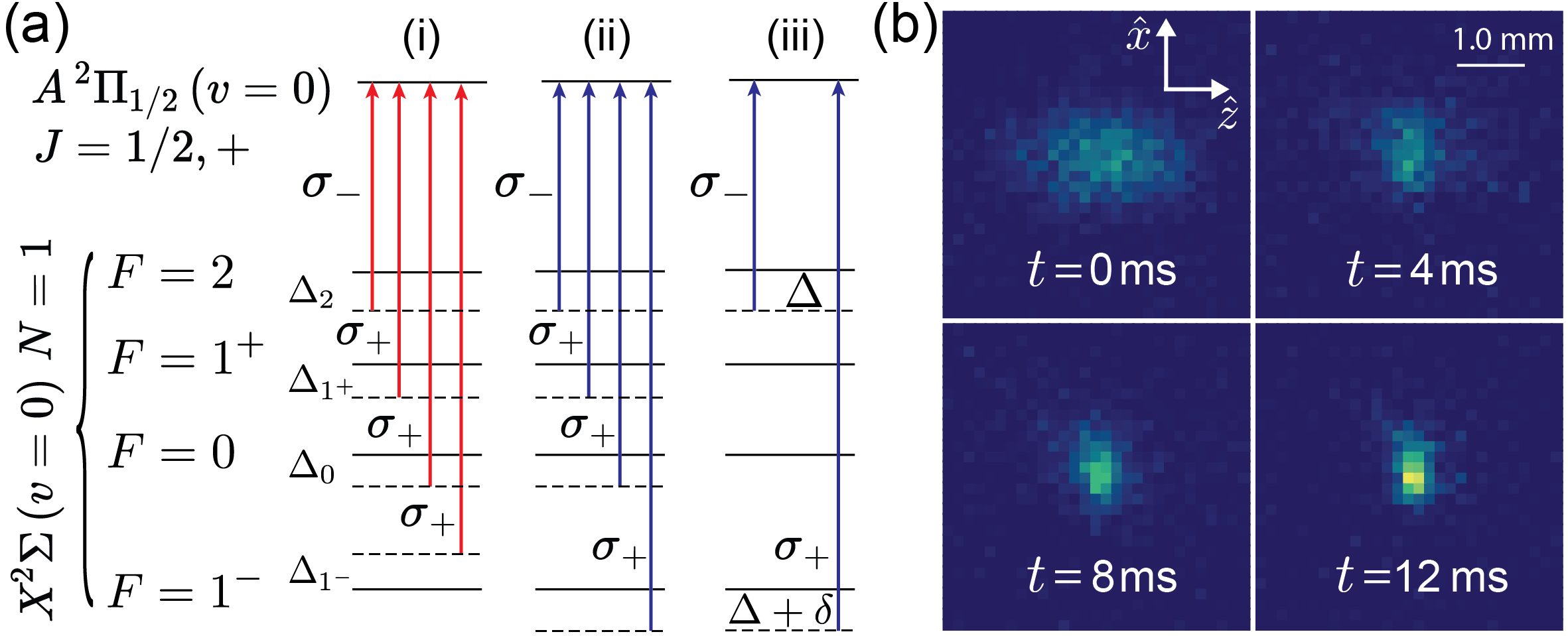}}
	\caption{\label{Fig_1} MOT Laser Configurations and $\Lambda$-BDM Loading Dynamics. (a) Laser configurations for (i) a conventional red-detuned DC MOT, (ii) the four-frequency BDM, and (iii) the two-frequency $\Lambda$-BDM. The single-photon (two-photon) frequency detuning is denoted $\Delta$ ($\delta$). The specific detunings and intensities are provided in~\cite{Supplement}. (b) In-situ images of a $\Lambda$-BDM at time $t$ following direct loading from a $\Lambda$-cooled cloud ($B_z'=27\,\text{G/cm}$).}
	\vspace{-0.2in}
\end{figure}

In this work, we demonstrate a BDM of CaF molecules, producing the coldest and densest CaF MOT reported to date. Our work starts with a conventional red-detuned MOT of CaF loaded from a cryogenic buffer gas beam. The red MOT utilizes a DC quadrupolar magnetic field (symmetry axis along $\hat{z}$) and laser-cooling light red-detuned from the $X\,^2\Sigma(N=1) \rightarrow A\,^2\Pi_{1/2}(J=1/2,+)$ transition. The light is sent along three orthogonal directions and consists of four frequency components nominally addressing the four ground hyperfine manifolds. The polarizations and frequency detunings of the components are shown in Fig.~\ref{Fig_1}(a,i). Initially, the red-detuned MOT captures molecules at an axial magnetic gradient $B_z' = 26\,\text{G/cm}$. Subsequently, the MOT is compressed by ramping the gradient to $B_z' = 104\,\text{G/cm}$~\cite{Williams2018MagTrap,Lu2021Ring} while simultaneously ramping down the light intensity. This produces a sample with a Gaussian size of $\sigma = 680(30)\,\mu\text{m}$ and a temperature of $T = 3.9(4)\,\text{mK}$.

 Since BDMs can have low capture velocities~\cite{devlin2016coolingSim,tarbutt2018blueMOT,Langin2023simulation,Xu2022blueMOTsim}, we perform further cooling~\cite{Burau2023BlueMOT,Jorapur2023blueMOT}. We switch off the magnetic gradient and immediately perform $\Lambda$-enhanced gray molasses cooling~\cite{Cheuk2018Lambda,Burau2023BlueMOT,Jorapur2023blueMOT} to reach a temperature of $\approx 10\,\mu \text{K}$. The sample expands negligibly over the $\sim\text{ms}$ cooling timescale, and $49(1)\%$ of the initial molecules remain.

Subsequently, molecules are captured into the BDM, which uses light blue-detuned to the $X\,^2\Sigma(N=1) \rightarrow A\,^2\Pi_{1/2}(J=1/2,+)$ transition. Empirically, we have found two BDM schemes: one with four frequencies addressing all four ground hyperfine manifolds, and another with two frequencies (Fig.~\ref{Fig_1}(a,ii,iii)). The first scheme is used for initial loading; the second scheme, which we call $\Lambda$-BDM, is used to produce the densest and coldest samples.

Initially, we transfer molecules into the four-frequency BDM with a magnetic gradient of $B_z'=14.6\,\text{G/cm}$ and an overall frequency detuning of $\approx 20\,\text{MHz}$ (details in~\cite{Supplement}). At optimal parameters, the transfer efficiency from the $\Lambda$-cooled cloud is $\approx 70\%$. Notably, the BDM rapidly reaches its equilibrium size over a $1/e$ time of $7.9(4)\,\text{ms}$. At equilibrium, the axial (radial) size is $\sigma_z = 294(2)\,\mu\text{m}$ ($\sigma_r = 215(2)\,\mu\text{m}$), and the axial (radial) temperature is $T_z= 195(5)\,\mu\text{K}$ ($T_r=186(12)\,\mu\text{K}$). The corresponding mean size is $\sigma= \sigma_r^{2/3} \sigma_z^{1/3} = 239(2)\,\mu\text{m}$ and the mean temperature is $T=T_r^{2/3} T_z^{1/3}=189(8)\,\mu\text{K}$, around the Doppler limit of $T_D= 200\,\mu\text{K}$. The equilibration time is similar to that reported for SrF and is significantly faster than observed in YO ($\sim50\,\text{ms}$). We attribute the rapid equilibration to the high photon scattering rate, which we measure to be $\Gamma_{\text{sc}}=2.0(2)\times 10^{6}~\text{s}^{-1}$~\cite{Supplement}. Although this high scattering rate could provide large trapping and damping forces, it limits the achievable temperature.

To reach lower temperatures, we subsequently switch to the two-frequency $\Lambda$-BDM configuration (Fig.~\ref{Fig_1}(a,iii)) with light addressing the highest ($J=3/2, F=2$) and lowest ($J=1/2, F=1^-$) hyperfine manifolds. This configuration is similar to that used in $\Lambda$-cooling, where velocity-dependent coherent dark states enable cooling to sub-Doppler temperatures. Similar to the four-frequency BDM, we also observe that the $\Lambda$-BDM rapidly approaches its equilibrium size over several milliseconds (Fig.~\ref{Fig_1}(b)).

\begin{figure}[t]
	{\includegraphics[width=\columnwidth]{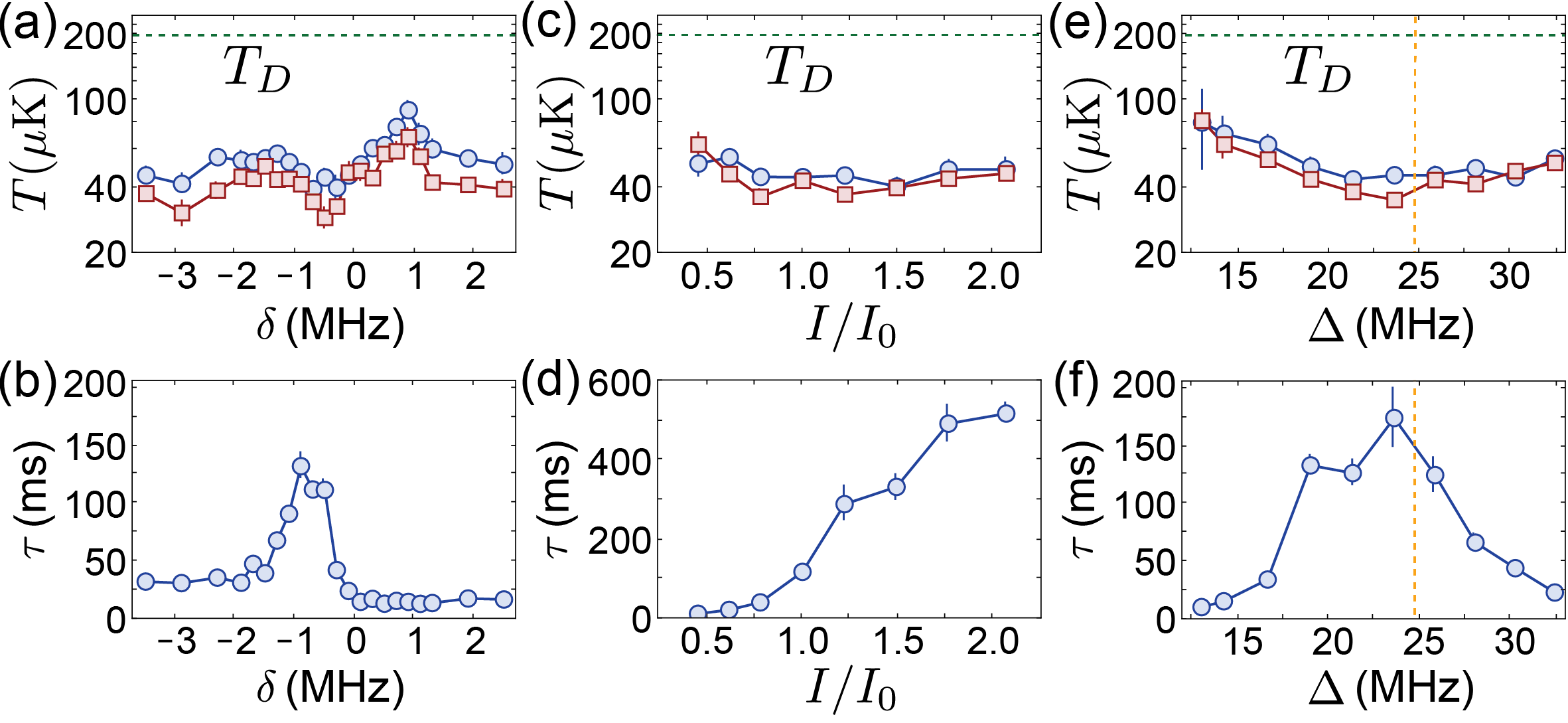}}
       \caption{\label{Fig_2} Parameter Dependences of the $\Lambda$-BDM Temperature $T$ and Lifetime $\tau$. (a),(b) $T$ and $\tau$ versus two-photon detuning $\delta$, with $\Delta = 23.8\,\text{MHz}$ and $I = 5.8(2)\,\text{mW/cm}^2$. (c),(d) $T$ and $\tau$ versus the intensity $I/I_0$ per beam ($I_0 = 5.8(2)\,\text{mW/cm}^2$, $\Delta = 23.8\,\text{MHz}$, $\delta=-0.7\,\text{MHz}$). (e),(f) $T$ and $\tau$ versus single-photon detuning $\Delta$ ($\delta=-0.7\,\text{MHz}$, $I = 5.8(2)\,\text{mW/cm}^2$). The vertical dashed line shows the value of $\Delta$ where the $F=2$ frequency component is resonant with the $F=1^+$ manifold. In (a),(c),(e), axial (radial) temperatures $T_z$ ($T_r$) are shown in blue circles (red squares) and the horizontal dashed line shows the Doppler temperature $T_D$. For all plots, $B'_z=10.4\,\text{G/cm}$.}
       \vspace{-0.2in}
\end{figure}

To characterize the $\Lambda$-BDM, we investigate the dependence of its temperature ($T$) and lifetime ($\tau$) on the single-photon frequency detuning ($\Delta$), two-photon frequency detuning ($\delta$), and intensity ($I$). We keep the gradient at $B_z'=10.4\,\text{G/cm}$ and first probe the dependences of $T$ and $\tau$ on $\delta$. As shown in Fig.~\ref{Fig_2}(a,b), we observe a temperature minimum and lifetime maximum near two-photon resonance ($\delta=0$), with a striking rise in temperature and decrease in lifetime for $\delta>0$. These features are similar to those observed in free-space $\Lambda$-cooling~\cite{Cheuk2018Lambda} and in a BDM of YO~\cite{Burau2023BlueMOT}, supporting the existence of velocity-dependent coherent dark states in the $\Lambda$-BDM. Notably, at $B_z'=10.4\,\text{G/cm}$ and $\delta = -0.7\,\text{MHz}$, we observe a mean temperature of $T = 36.7(2)\,\mu\text{K}$, a factor of 5 below the Doppler limit of $T_D=200\,\mu\text{K}$~(Fig.~\ref{Fig_2}(a)). We have verified that the $\Lambda$ features, a minimum in $T$ and a maximum in $\tau$ versus $\delta$, persist at a higher gradient of $33.6\,\text{G/cm}$~\cite{Supplement}.

We next explore the dependences of $T$ and $\tau$ on beam intensity $I$. Similar to free-space $\Lambda$-cooling~\cite{Cheuk2018Lambda}, the robustness of the dark states increases with optical intensity due to increased two-photon coupling. Because the two-photon resonance varies across the BDM due to the magnetic gradient, one expects that a minimum intensity is needed to counteract this magnetic broadening. However, at high intensities, photon scattering is increased and could lead to higher temperatures. We thus expect an optimal intensity to exist. We indeed observe an optimal intensity that produces a minimum temperature (Fig.~\ref{Fig_2}(c,d)). Notably, the lifetime improves with intensity at the expense of higher temperatures. As a compromise, we choose to operate at $I = I_{0}=5.8(2)\,\text{mW/cm}^2$.

Similar to the dependence on $I$, we expect an optimum in the single-photon detuning $\Delta$. The optimum occurs as a balance between increased off-resonant scattering at small $\Delta$ and decreased two-photon coupling at large $\Delta$, which reduces the robustness of coherent dark states~\cite{Cheuk2018Lambda}. At a fixed intensity of $I= 5.8(2)\,\text{mW/cm}^2$ and two-photon detuning of $\delta= -0.7\,\text{MHz}$, we find an optimum in lifetime and temperature at $\Delta \approx 24\,\text{MHz}$ (Fig.~\ref{Fig_2}(e,f)). At the optimal parameters ($\Delta =23.8\,\text{MHz}$, $\delta=-0.7\,\text{MHz}$, $I= 5.8(2)\,\text{mW/cm}^2$), we measure a scattering rate of $\Gamma_{\text{sc}}=0.65(19)\times10^6\,\text{s}^{-1}$~\cite{Supplement}, substantially lower than the four-frequency BDM, but much higher than free-space $\Lambda$-cooling~\cite{Cheuk2018Lambda}. 

We next explore the dependence of the $\Lambda$-BDM on magnetic gradient $B_z'$. This investigation is motivated by the practical objective of increasing molecular densities. A viable strategy in red-detuned MOTs is to compress molecular samples by increasing $B_z'$~\cite{Williams2017MOTchar}. At a fixed temperature, the MOT size decreases with gradient as $1/\sqrt{B_z'}$, and the density consequently rises as $B_z'\,^{3/2}$~\cite{Williams2017MOTchar,Supplement}. In detail, the MOT can be modeled with an equation of motion that captures a restoring force and velocity damping: $\ddot{x} = -\alpha(x) \dot{x} + F(x)/m$~\cite{barry2014SrFMOT,Anderegg2017MOT,Williams2017MOTchar,Burau2023BlueMOT}. Near the MOT center and at low velocities, the damping coefficient $\alpha$ is approximately constant, and the restoring force is approximated by Hooke's law $F(x) = -kx$, where $k$ is the spring constant. The restoring force arises from magnetic-field dependent light scattering. Therefore, one can express the restoring force $F$ as a function of the local magnetic field $B$. At the center of a quadrupole field used in a MOT, $F(x) = F(B(x)) \approx B' x\,\left( dF/dB\right)$. Hence, near the center of the MOT where Hooke's law is valid, $k \propto B_z'$. Using a generalized Virial theorem that equates potential and kinetic energy~\cite{Supplement}, one finds that $\sigma \propto 1/\sqrt{B_z'}$ and the density grows as $(B_z')^{3/2}$ at a fixed temperature $T$.

\begin{figure}[t]
	{\includegraphics[width=\columnwidth]{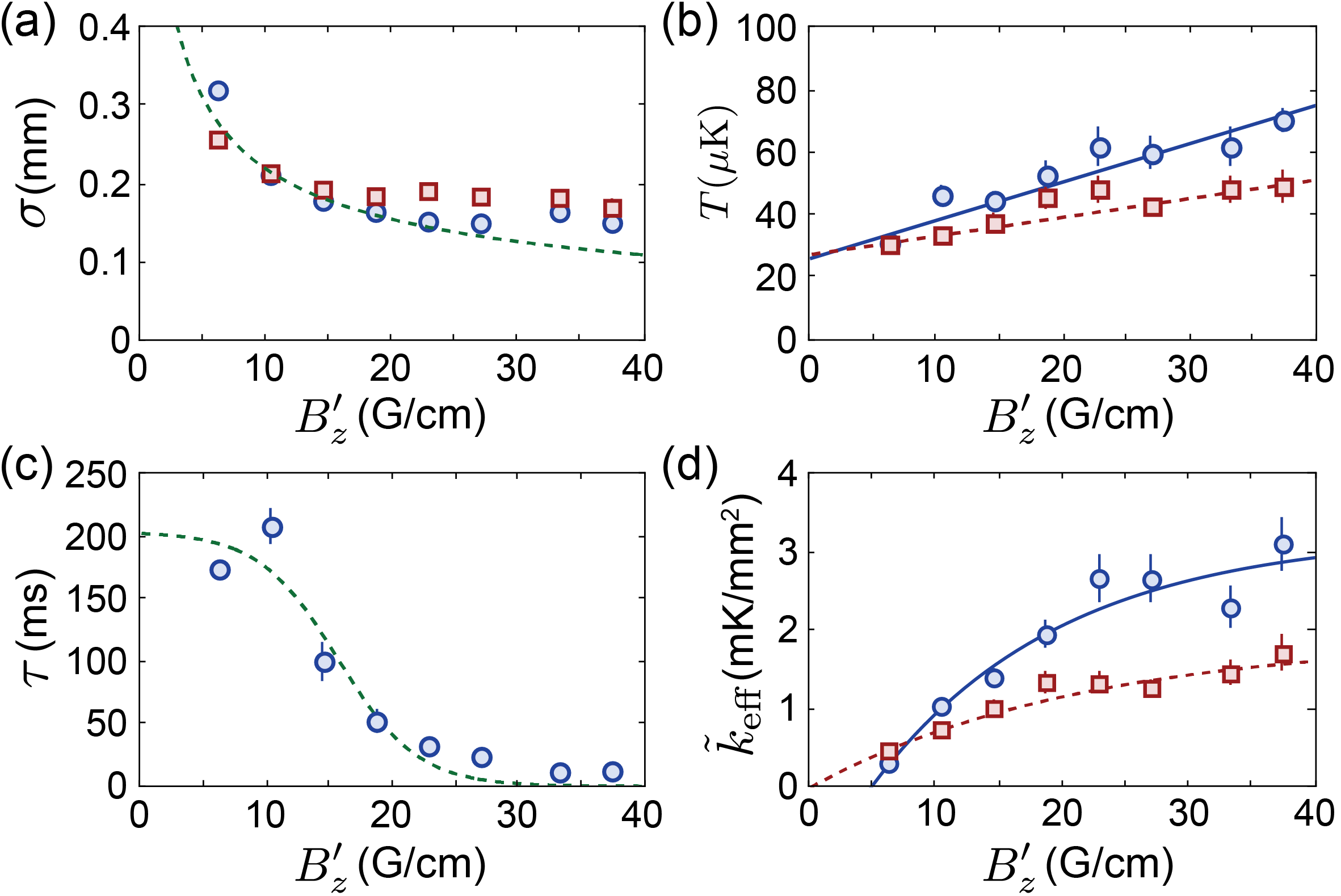}}
        \caption{\label{Fig_3} $\Lambda$-BDM Properties versus Axial Magnetic Gradient $B_z'$. (a) Gaussian axial (radial) width $\sigma_z$ ($\sigma_r$) shown in blue circles (red squares). Green dashed line shows a fit to the mean size $\sigma= \sigma_r^{2/3} \sigma_z^{1/3}$ to $1/B_z^{1/2}$ for data up to $20\,\text{G/cm}$. (b) Axial (radial) temperature $T_z$ ($T_r$) shown in blue circles (red squares). The blue solid (red dashed) line is a linear fit to the axial (radial) temperature. (c) Lifetime $\tau$ shown in blue circles. Dashed line is a fit to a logistic function that serves as a guide to the eye. (d) Effective spring constant $\tilde{k}_{\text{eff}} = T/\sigma^2$ for the axial (radial) direction shown in blue circles (red squares). The blue solid (red dashed) line is a fit of the axial (radial) data to an exponential saturation curve. The axial fit includes a horizontal offset to phenomenologically capture the effect of gravity, which is along $\hat{z}$.}
        \vspace{-0.2in}
\end{figure}

In Fig.~\ref{Fig_3}, we show the observed dependence of the size ($\sigma$), temperature ($T$), and lifetime ($\tau$) on $B_z'$. We find that $\sigma$ follows the scaling of $1/\sqrt{B_z'}$ up to $\approx 20\,\text{G/cm}$ (Fig.~\ref{Fig_3}(a)). $T$ increases with gradient, but remains well below $T_D$ even up to $40\,\text{G/cm}$ (Fig.~\ref{Fig_3}(b)). The rising temperature can be explained by the fact that molecules experience a mean magnetic field that rises with gradient as $\sqrt{B_z'}$ at fixed $T$~\cite{Supplement}, which in turn perturbs the coherent dark states responsible for the $\Lambda$-cooling features. We also find that $\tau$ decreases significantly starting at $B_z' \approx 15\,\text{G/cm}$ (Fig.~\ref{Fig_3}(c)). At $30\,\text{G/cm}$, $\tau$ becomes comparable to the equilibration timescale. This renders density enhancement via magnetic compression of the BDM impractical. We note that the lifetime can be increased using higher intensities at the expense of higher temperatures~\cite{Supplement}.

The optimal gradient is therefore a compromise between minimizing $\sigma$ and $T$ and maximizing $\tau$. Empirically, the smallest sizes occur at $B_z'=27\,\text{G/cm}$ ($\sigma=172(4)\,\mu\text{m}$, $T=60(4)\,\mu\text{K}$), the lowest temperatures at $B_z'=6.2\,\text{G/cm}$ ($\sigma=277(5)\,\mu\text{m}$, $T=31(1)\,\mu\text{K}$), and the highest peak phase space density (PSD) at $B_z'=14.6\,\text{G/cm}$ ($\sigma=188(3)\,\mu\text{m}$, $T=39(2)\,\mu\text{K}$). At optimal parameters and with $N=6.2(15) \times 10^3$ molecules in the $\Lambda$-BDM, we obtain a peak density of $n_0=7(2)\times10^7\,\text{cm}^{-3}$ and a peak PSD of $3.0(8)\times10^{-9}$. Compared to the compressed red-detuned MOT, these values correspond to a density improvement of 19(3) and a PSD enhancement of $1.6(3) \times10^4$. Notably, by applying free-space $\Lambda$-cooling to a cloud released from a $\Lambda$-BDM ($B_z'=18.7\,\text{G/cm}$), we obtain a peak PSD of $2.3(6) \times 10^{-8}$ ($\sigma=182(2)\,\mu\text{m}$, $T=10.6(6)\,\mu\text{K}$), the highest reported to date for CaF in free space.

To investigate the origin of the short lifetimes at high gradients, we first examine the dependence of the restoring force on $B_z'$, since an insufficient restoring force could lead to loss. Assuming Hooke's law and constant velocity damping, the spring constant $k$ can be obtained from $T$ and $\sigma$ via $k_{\text{eff}}=k_B T/\sigma^2$~\cite{Supplement}. In Fig.~\ref{Fig_3}(d), we show $\tilde{k}_{\text{eff}}= k_{\text{eff}}/k_B$ versus $B_z'$ for both the radial and axial directions. At gradients below $B_z'\approx 10\,\text{G/cm}$, $\tilde{k}_{\text{eff}}$ is linear in $B_z'$, as expected, and has values similar to those reported for YO~\cite{Burau2023BlueMOT}. At higher gradients, $\tilde{k}_{\text{eff}}$ appears to saturate, indicating that the average restoring force decreases. Because molecules experience an average magnetic field that grows with gradient as $\sqrt{B_z'}$, the saturation of $\tilde{k}_{\text{eff}}$ suggests that $F(B)$ is significantly sub-linear in $B$ at higher magnetic fields.

To test this hypothesis, we next directly measure $F(B)$. We create dense and cold samples by loading a $\Lambda$-BDM and subsequently applying free-space $\Lambda$-cooling. We then impart an initial velocity of $v_0 \approx 200\,\text{mm/s}$ in the radial direction by pulsing on a beam resonant with the $X{}^2\Sigma(N=1) \to B{}^2\Sigma(N=0)$ transition. We next apply a uniform magnetic field $\vec{B}$ along the push direction ($\vec{B}\parallel\vec{v}_0$) and switch on the $\Lambda$-BDM light for a variable duration $t$. For each $t$, we measure the velocity $v$ via time-of-flight expansion. By fitting to an exponential decay with an offset, $v(t) = A e^{-\alpha t} + v_{\infty}$, we extract the damping coefficient $\alpha$ and the terminal velocity $v_{\infty}$. The restoring force $F(B)$ is then obtained via $F= m\, \alpha \, v_{\infty}$. To determine $v_{\infty}$ more accurately, we perform a second set of measurements where the push beam is not applied and the molecules start with $v_0=0$ (see~\cite{Supplement} for details).

In Fig.~\ref{Fig_4}(a,b), we show the extracted damping curve $\alpha(B)$ and acceleration curve $a(B) = F(B)/m$. For $a(B)$, we observe the expected sign change when reversing the magnetic field. We also find that $\alpha(B)$ and $a(B)$ are significant only when $|B|<1\,\text{G}$. Furthermore, the restoring force is linear only for $|B|<0.5\,\text{G}$. At a gradient of $B_z'=20\,\text{G/cm}$, this magnetic field range corresponds to a radial size of $0.5\,\text{mm}$.

\begin{figure}
	{\includegraphics[width=\columnwidth]{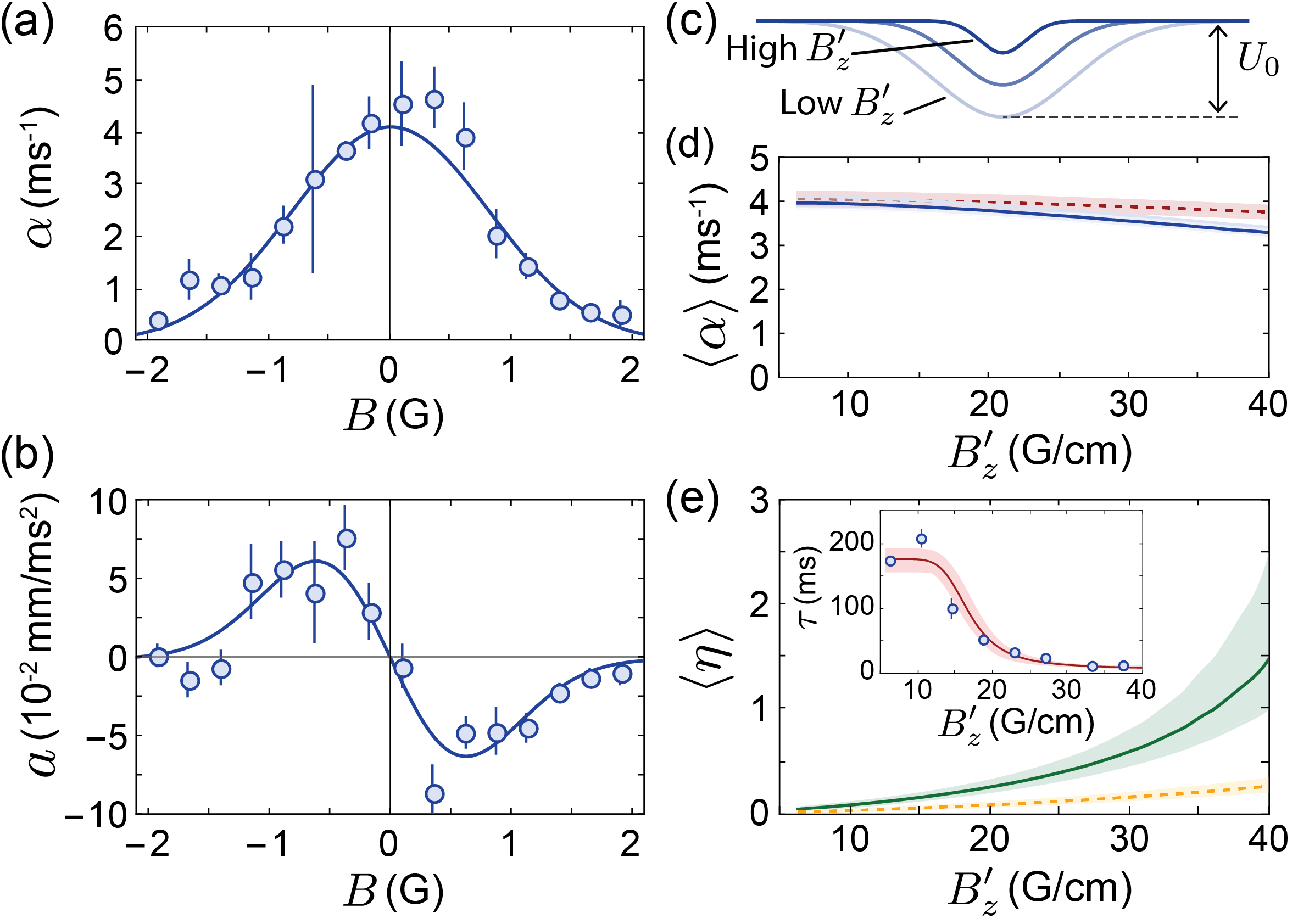}}
      \caption{\label{Fig_4} Velocity Damping, Restoring Force, and Effective Trap Depth of the $\Lambda$-BDM. (a) Damping constant $\alpha$ versus magnetic field $B$. The solid blue line is a Gaussian fit. (b) Acceleration $a(B)$ versus $B$. Solid line is a fit to the derivative of a Gaussian. (c) Illustration showing decreasing trap depths $U_{\text{0}}$ and trap sizes $d$ with increasing magnetic gradient $B_z'$. Both $U_{\text{0}}$ and $d$ scale as $1/B_z'$. (d) Average damping $\langle \alpha \rangle$ versus $B_z'$, simulated using the measured dependences  ($T(B_z')$, $\sigma(B_z')$ and $\alpha(B)$.) $\langle \alpha \rangle$ along the axial (radial) directions are shown by the blue solid line (red dashed line). (e) Average temperature-to-trap-depth ratio $\langle \eta \rangle$ along the axial (radial) directions are shown by the green solid line (orange dashed line). Inset: measured lifetime $\tau$ versus $B_z'$ shown in blue circles, with the solid line a fit to the finite trap depth loss model described in the main text. For (d),(e), shaded regions indicate the uncertainty bands obtained using $1\sigma$ uncertainties in the fitted parameters for $\alpha(B)$, $a(B)$, $T(B_z')$ and $\sigma(B_z)$~\cite{Supplement}. }
      \vspace{-0.2in}
\end{figure}

The acceleration curve $a(B)$ provides an explanation of the short lifetimes observed at high gradients. In short, we find that the effective BDM trap depth becomes comparable to the molecular temperatures at high gradients, leading to loss. In detail, using $a(B)$, we define an effective conservative trapping potential $U(x) = - m \int a(B(x)) dx$ along the radial and axial directions~\cite{ConsvFnote}. It follows that both the spatial scale and the magnitude of $U(x)$ scale inversely with $B_z'$ (Fig.~\ref{Fig_4}(c)). In other words, both the size and depth of the effective trap decrease with $B_z'$. To extract an effective depth, we assume that $U(x)$ is of a Gaussian form given by $U(x) = -U_{\text{0}} \exp{[-x^2/(2d^2)]}$, where $U_{\text{0}}$ is the maximum trap depth and $d$ the Gaussian spatial width. Fitting $a(B)$, we obtain a trap depth of $U_{\text{0}}(B_z') = k_B \times 4.6(10) \, \text{mK}\cdot \text{(G/cm)}/ B_z'$, and a trap size of $d(B_z')= 0.62(4)\,\text{G} /B_z'$. Notably, at a typical gradient of $20\,\text{G/cm}$, $U_{\text{0}} = k_B \times 230(50)\,\mu\text{K}$, which is only four times higher than the observed temperature. These trap depths are much smaller than the $\sim 50\,\text{mK}$ depths reported for red-detuned molecular MOTs~\cite{McCarron2015improvedMOT,Williams2017MOTchar,Langin2023simulation} and $\sim 0.5\,\text{K}$ depths reported for red-detuned atomic MOTs~\cite{Raab1987NaMOT}.

To further support our hypothesis that the short lifetimes arise from insufficient trap depths, we quantitatively examine both the damping $\alpha$ and the local temperature-to-trap-depth ratio $\eta(x) = (k_B T)/U(x)$. Using the measured dependences of $\sigma$ and $T$ on $B_z'$ (Fig.~\ref{Fig_3}), we perform Monte-Carlo simulations to obtain ensemble-averaged values of $\bra \eta \ket$ and $\bra \alpha \ket$ (Fig.~\ref{Fig_4}(d,e))~\cite{Supplement}. We observe that $\bra \alpha \ket$, which determines the average cooling rate, varies by no more than $20\%$ over experimentally relevant gradients. On the other hand, $\bra \eta \ket$ varies significantly and even exceeds unity. Because the fraction of particles above the local trap depth is given by $\exp(-1/\eta)$, the dependence of $\bra \eta \ket$ on $B_z'$ suggests that a shallow trap depth (insufficient restoring force), rather than the lack of velocity damping, is the primary cause of loss at high gradients. Semi-quantitatively, the loss rate due to finite trap depth can be estimated as $ C \sum_i \langle \alpha_i \exp(-1/\eta_i) \rangle$, where $i$ sums over the radial and axial directions and $C$ is a constant of order $\mathcal{O}(0.1)$~\cite{Supplement}. As shown in the inset of Fig.~\ref{Fig_4}(e), this loss model (with two additional parameters to describe background loss and minimum escape time) qualitatively reproduces the observed lifetime dependence on $B_z'$.

In summary, we have demonstrated a blue-detuned MOT of CaF molecules that reaches temperatures significantly below the Doppler limit. Specifically, the $\Lambda$-BDM provides the smallest sizes, lowest temperatures, and highest phase space densities reported to date for magneto-optically trapped CaF molecules. We have also uncovered a mechanism that hinders further density enhancement via magnetic compression. We find low effective $\Lambda$-BDM trap depths that decrease with the strength of the magnetic gradient, leading to short lifetimes at high gradients. Looking ahead, BDMs could significantly aid the loading of molecules into optical tweezer arrays, which are a promising platform for quantum simulation and quantum information processing~\cite{demille2002quantum,Carr2009review,Bohn2017molreview,blackmore2018reviewMoleculeQuantum,Holland2022Tweezer,Bao2022Tweezer}. In addition, the enhanced densities could also aid the exploration of ultracold molecular collisions~\cite{Cheuk2020collisions,Jorapur2023blueMOT} or provide better starting conditions for evaporative cooling.

We conclude by contextualizing our BDM with those recently reported for YO~\cite{Burau2023BlueMOT} and SrF~\cite{Jorapur2023blueMOT}. Despite different spin-rotation and hyperfine structure, we find that a CaF BDM is also possible. Specifically, the $\Lambda$-BDM robustly achieves sub-Doppler temperatures and substantially higher densities compared to a red MOT. This suggests that BDMs could be a powerful technique widely applicable to many laser-coolable molecules, including polyatomic ones such as CaOH and SrOH~\cite{baum20201d,Vilas2022MOTCaOH,kozyryev2017sisyphus,mitra2020CaOCH3}.

\begin{acknowledgments}
We thank Callum Welsh for a careful reading of the manuscript. This work is supported by the National Science Foundation under Grant No. 2207518. S.J.L. acknowledges fellowship support from a Princeton Quantum Initiative Graduate Student Fellowship. C.M.H. acknowledges support from a Joseph Taylor Graduate Student Fellowship. L.W.C. acknowledges support from the Sloan Foundation.
\end{acknowledgments}

\bibliographystyle{apsrev4-1}

\begin{thebibliography}{38}
\bibitem [{Con()}]{ConsvFnote}%
  \BibitemOpen
  \href@noop {} {}\bibinfo {note} {The restoring force is in general
  non-conservative, but it is conservative when restricted to the $x$-$y$ plane
  ($z=0$) or along the symmetry axis $z$.}\BibitemShut {Stop}%
\end{thebibliography}

\end{document}


\title{Supplementary Information for ``A Blue-Detuned Magneto-Optical Trap of CaF Molecules" }

\author{Samuel J. Li}
\thanks{These authors contributed equally to this work.}
\affiliation{Department of Physics, Princeton University, Princeton, New Jersey 08544 USA}
\author{Connor M. Holland}
\thanks{These authors contributed equally to this work.}
\affiliation{Department of Physics, Princeton University, Princeton, New Jersey 08544 USA}
\author{Yukai Lu}
\affiliation{Department of Physics, Princeton University, Princeton, New Jersey 08544 USA}
\affiliation{Department of Electrical and Computer Engineering, Princeton University, Princeton, New Jersey 08544 USA}
\author{Lawrence W. Cheuk}
\email{lcheuk@princeton.edu}
\affiliation{Department of Physics, Princeton University, Princeton, New Jersey 08544 USA}

\maketitle

\section{Methods}
\subsection{Preparation of Molecules}
CaF molecules are produced in a single-stage cryogenic buffer gas cell~\cite{Hutzler2012CBGB}.
Molecules in the $X{}^2\Sigma(v=0, N=1)$ manifold
are slowed via chirped slowing and loaded into a red-detuned DC magneto-optical trap (MOT) operating on the $X{}^2\Sigma(v=0,N=1) \to A{}^2\Pi_{1/2}(v=0, J=1/2,+)$ transition ($\lambda=606.3\,\text{nm}$)~\cite{holland2022bichromatic}.

The MOT light contains four frequencies addressing the four ground hyperfine levels in $X{}^2\Sigma(v=0,N=1)$: $\ket{J,F} = \ket{3/2, 2}$, $\ket{3/2, 1}$, $\ket{1/2, 0}$ and $\ket{1/2, 1}$. As a shorthand, we will refer to the hyperfine levels by $F$ and disambiguate $\ket{3/2, 1}$ and  $\ket{1/2, 1}$ by labeling them $F=1^+$ and $F=1^-$, respectively.
The excited hyperfine structure of the $A{}^2\Pi_{1/2}(v=0, J=1/2,+)$ is unresolved. The four frequency components have single-photon frequency detunings of $\Delta+  ( - 2.3\,\text{MHz}, 0, 0,  - 9.6\,\text{MHz})$, with $\Delta= 0.7\,\text{MHz}$. The single-photon frequency detunings are given with respect to the transitions to the excited hyperfine state of $\left|A{}^2\Pi_{1/2}(v=0, J=1/2),F=1,m_F=0\right\rangle$. The relative power ratios are 5:3:1:3. The handedness of the $F=2$ component is opposite to the three other ones.

\begin{figure}[t]
	\includegraphics[width=\columnwidth]{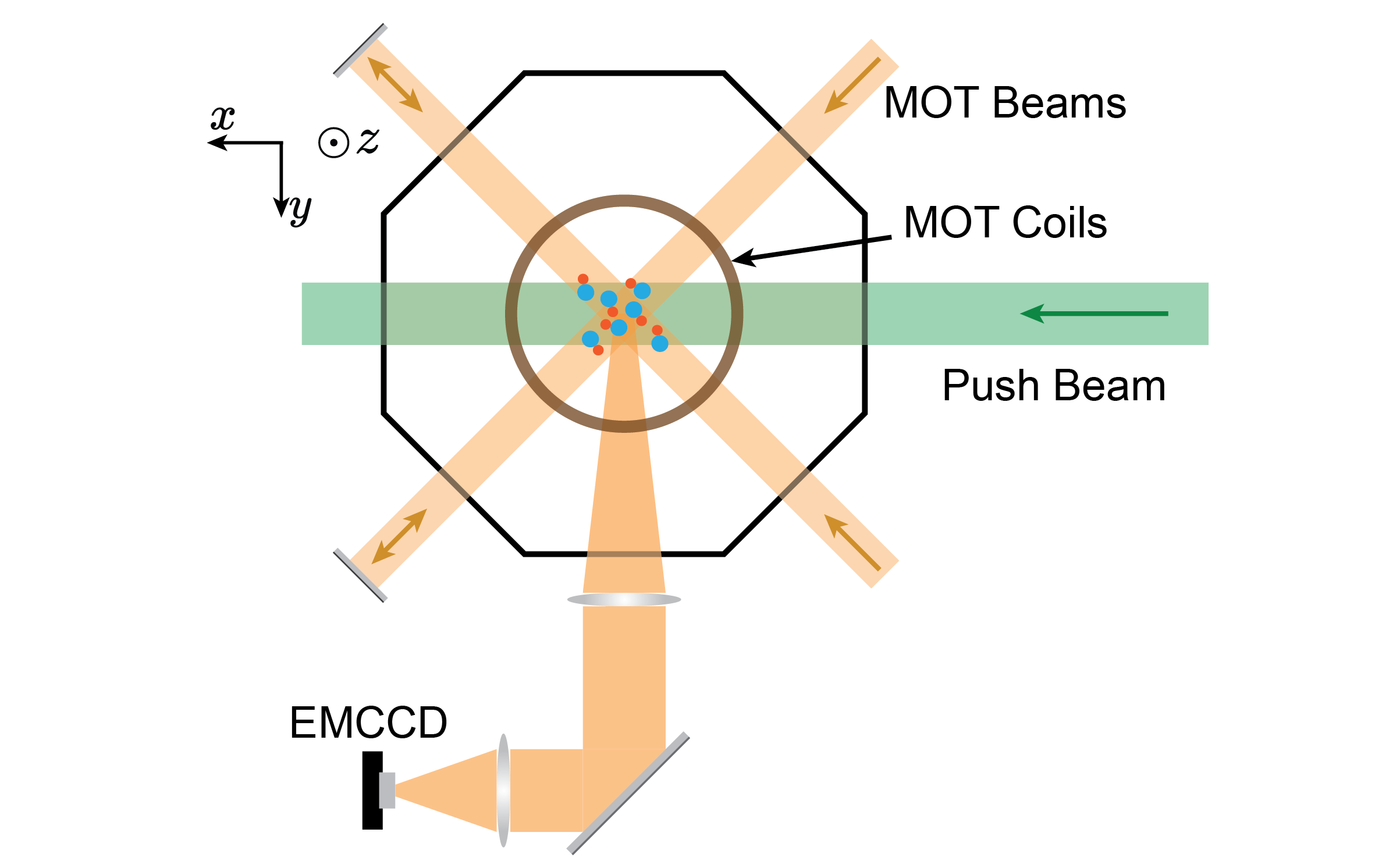}
	\caption{\label{fig: beam diagram} Beam Configuration. Six MOT beams containing light detuned from the $X-A$ transition ($\lambda=606.3\,\text{nm}$) propagate along three orthogonal axes. A magnetic field gradient is applied using a pair of coils in near anti-Helmholtz configuration. Three additional pairs of coils in near-Helmholtz geometry allow independent control of a bias magnetic field. A beam containing light resonant with the $X-B$ transition ($\lambda = 531.0\,\text{nm}$) propagating along $\hat{x}$ is used as a ``push beam'' to impart an initial velocity to the molecules. Molecules are imaged by collecting $X-A$ fluorescence along the $y$ direction using an EMCCD camera. The gravitational force is along the $z$-axis.}
\end{figure}

The MOT light is sent along three orthogonal directions ($x,y,z$). The beams in the horizontal plane ($xy$-plane) are retro-reflected, while the two beams along the vertical direction ($z$) are sent in separately (Fig. \ref{fig: beam diagram}). Each horizontal MOT beam has a total power of $\sim 22\,\text{mW}$ with a beam waist ($1/e^2$ radius) of $\sim6\,\text{mm}$. The intensity of each vertical beam is matched to that of the radial beams to within $20\%$, and each vertical beam has a beam waist of $\sim 2.5\,\text{mm}$. 

The magnetic field gradient is generated using a pair of anti-Helmholtz coils, whose symmetry axis is along the $z$ direction. This implies that the axial magnetic gradient $B_z' = d B_z/dz$ is related to the radial magnetic gradient $B_r' = dB_r/dr$ via $B_z'=2B_r'$. The initial loading of molecules into the DC red-detuned MOT occurs for 5\,ms at an axial magnetic gradient of $B_z' = 26\,\text{G/cm}$. Subsequently, to compress the MOT, the gradient is linearly increased to $B_z'=104\,\text{G/cm}$ and the MOT optical power is decreased by a factor of $4.5$ over $10\,\text{ms}$. The compression reduces the $1/e^2$ radius from $2\sigma \sim 2\,\text{mm}$ to $1.0\,\text{mm}$. Details of the compression procedure can be found in~\cite{Lu2021Ring}.

Unless otherwise noted, all MOT and imaging beams throughout this work contain vibrational repumping light addressing the $X{}^2\Sigma(v, N=1) \to A{}^2\Pi_{1/2}(v-1, J=1/2,+)$ transition ($\lambda \approx 628\,\text{nm}$) for $v = 1$. Vibrational repumps for $v=2,3$ are independently delivered along $\hat{x}$. Rotational repumping light addressing the $X{}^2\Sigma(v=0, N=3) \to B{}^2\Sigma(v=0, N=2)$ transition ($\lambda = 531.0\,\text{nm}$) is delivered along the horizontal MOT beam paths. All repumpers can be independently switched on and off.

\subsection{BDM Configurations and Loading Sequences}
Following compression of the DC red-detuned MOT, we turn off the magnetic gradient and apply $\Lambda$-enhanced gray molasses ($\Lambda$-cooling)~\cite{Cheuk2018Lambda} to cool the molecules to $\sim10 \,\mu\text{K}$. Details concerning the specific $\Lambda$-cooling sequence can be found in~\cite{Lu2021Ring}. We find that 49(1)\% of molecules are transferred from the red MOT to the $\Lambda$-cooled cloud. This cooling step helps increase the transfer fraction into the blue MOT.

The procedure to transfer $\Lambda$-cooled molecules to the $\Lambda$-BDM consists of the following four stages.

1) We form a four-frequency blue-detuned MOT (BDM). The four frequency components nominally address the four hyperfine ground manifolds with single-photon frequency detunings $\Delta +  (2.2\,\text{MHz}, 0, 0,  4.2\,\text{MHz})$, where $\Delta=20.6\,\text{MHz}$. The power ratio of the four components is 5:1.5:1:3. The total power in each horizontal beam is $13.2(1)\,\text{mW}$; the vertical beams have similar intensities as the horizontal beams. The magnetic gradient is initialized at a low value and then held or increased to its final value over $5\,\text{ms}$ during this four-frequency BDM stage for compression. The values for the initial gradient and final gradients used in this work are listed in Tab.~\ref{tab:BGradShim}.

2) Following the four-frequency BDM, we switch to a two-frequency $\Lambda$-BDM configuration while keeping the magnetic gradient unchanged. The remaining frequencies address the $F=2$ and $F=1^-$ hyperfine manifolds, and have a relative power ratio of $1.4(1) : 1$. The single-photon frequency detuning $\Delta$ of the $F=2$ component is jumped to $23.8\,\text{MHz}$, and the two-photon frequency detuning is $\delta=-1.95\,\text{MHz}$. The total power is decreased so that each horizontal beam contains $8.0(1)\,\text{mW}$ of power. This configuration is held for $5\,\text{ms}$.

3) After an initial two-frequency BDM stage, the overall MOT beam power is decreased over $5\,\text{ms}$ to $3.5(1)\,\text{mW}$ in each beam. This configuration is held for $5\,\text{ms}$. 

4) Finally, we jump to the final two-frequency $\Lambda$-BDM configuration with $\delta=-0.75\,\text{MHz}$. The total power in each beam is jumped to $1.9(1)\,\text{mW}$. Depending on the specific data series, the parameters (single-photon detuning, two-photon detuning, and intensity) in this final stage are varied.

Unless otherwise noted, the peak intensity of each beam in the $\Lambda$-BDM configuration is $I = 5.8(2)\,\text{mW/cm}^2$.

Initially, we observe significant shearing and movement of the BDM as the magnetic gradient is increased during the BDM compression stage. This limits the lifetime of the BDM and lengthens the equilibration timescale. We believe that this is due to power imbalances between opposing MOT beams, which leads to a slight offset of the BDM center position from the location of zero magnetic field. To counteract this effect, we ramp an external bias magnetic field in rough proportion to $B_z'$ to ensure that the equilibrium BDM position remains largely fixed.

\begin{table}[h!]
	\begin{tabular}{c|c|c|c|c}
		$B_{z,i}'$ (G/cm) & $B_{z,f}'$ (G/cm)\\ 
		\hline
		6.2 & 6.2 \\ 
		10.4 & 10.4 \\ 
		14.6 & 14.6 \\ 
		14.6 & 18.7 \\
		14.6 & 22.9 \\ 
		14.6 & 27.0 \\ 
		14.6 & 33.3 \\ 
		14.6 & 37.4 \\ 
	\end{tabular}
	\caption{BDM Initial and Final Magnetic Field Gradients. To compress the four-frequency BDM, we ramp the $B$-field gradient from $B_{z,i}'$ to $B_{z,f}'$. The external bias magnetic field is adjusted to keep the MOT centered during the ramps. The gradient and bias fields are held at the specified values for all subsequent BDM stages.}
	\label{tab:BGradShim}
\end{table}

\begin{figure}
	\includegraphics[width=\columnwidth]{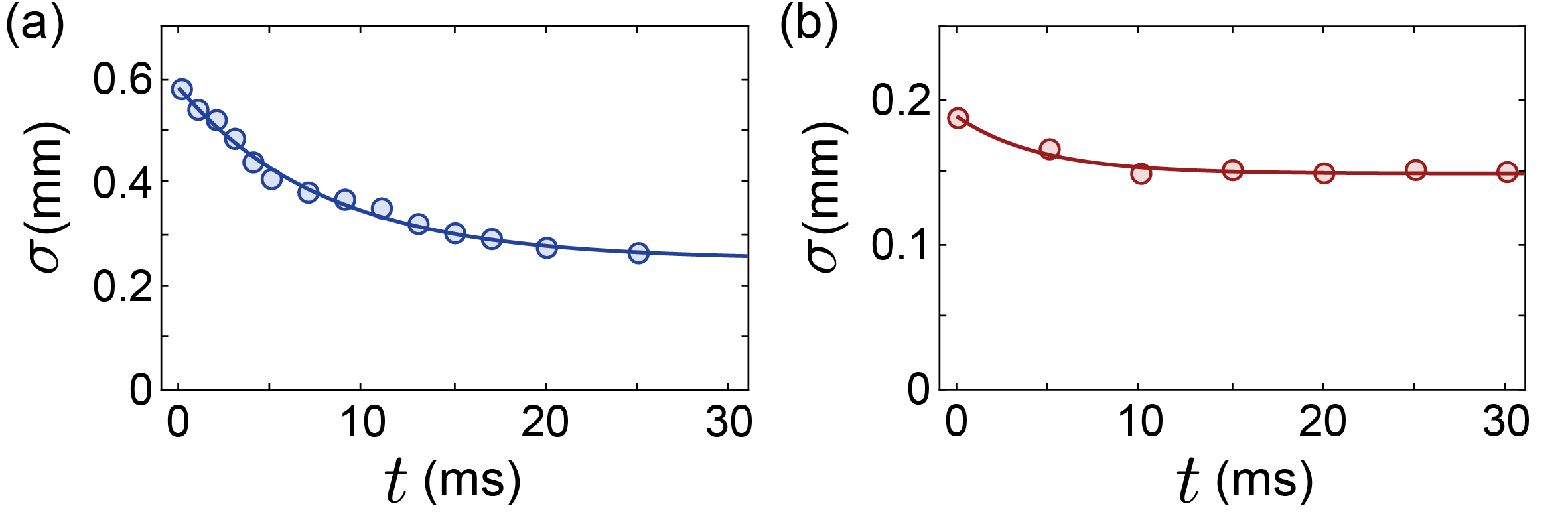}
	\caption{\label{fig:mot_timescale} In-situ Gaussian width $\sigma$ versus hold time $t$. (a) Mean width $\sigma$ of the four-frequency BDM ($B'_z=14.6\,\text{G/cm}$, $I=40(3)\,\text{mW/cm}^2$) versus $t$. The solid line shows a fit using an exponential curve with an offset. The fitted time constant is $7.9(4)\,\text{ms}$. (b) Mean width $\sigma$ of the two-frequency BDM (optimized for lifetime at $B'_z=33.3\,\text{G/cm}$) versus $t$. The fitted time constant is $4.5(11)\,\text{ms}$. For (a), $t=0$ is referenced to the beginning of stage 1. For (b), $t=0$ is referenced to the beginning of stage 4.}
\end{figure}

\section{Measurement Procedures}
\subsection{Measuring Molecular Cloud Size}
\label{sec:mot_size}
Molecules are imaged by exciting them on the $X\rightarrow A$ transition and collecting the resulting fluorescence on an EMCCD camera. The optical axis of the imaging system is along $\hat{y}$,
allowing measurement along the radial ($\hat{x}$) and axial ($\hat{z}$) dimensions of the cloud (Fig. \ref{fig: beam diagram}). Nearly all images are obtained as follows.

For each sequence, two images are taken under identical light conditions -- one containing molecules and one not containing molecules. This procedure allows us to isolate the molecular fluorescence by subtracting stray background light. Typically, 30-100 background-subtracted images are averaged. 

For data sets in which the center-of-mass or widths are measured, we integrate the averaged image along the two camera axes (aligned with $\hat{x}$ and $\hat{z}$) in an appropriate region of interest (ROI). We fit a Gaussian to the resulting signal. Using the known magnification of the imaging system, the cloud size and center-of-mass position along the $x$ and $z$ axes are then determined.

To obtain the in-situ images appearing in the main text Fig.~1(b) and the cloud size dynamics data in Fig.\ref{fig:mot_timescale}(a), we employ a four-image procedure. This procedure enables robust subtraction of a smearing effect caused by the non-zero shifting and cleaning speed of our EMCCD camera. In detail, we first take a pair of images of the molecular cloud (as described above) to generate the in-situ image. Next, we take a second pair of images during an identical experimental sequence, but turn off the imaging light during the two camera exposures to generate the ``smear" image. We next normalize the ``smear" image to a corresponding region in the in-situ image, and then subtract the two images.

\subsection{Measuring Temperature}
\label{sec:temperature}
The temperature of the molecular cloud is determined via a time-of-flight measurement.
The cloud is released from the MOT and expands freely for a variable time $t_m$ in the absence of light. It is then imaged using either $\Lambda$-cooling light or resonant light. For the data in Fig. 2(a) and Fig. 3(c), we image the cloud by applying $\Lambda$-cooling light for $30\,\text{ms}$. For Fig. 2(c) and (e), we resonantly image the cloud for $1\,\text{ms}$. A Gaussian fit is then applied to extract the sizes $\sigma_z$ and $\sigma_r$ as a function of $t_m$, along the two directions $\hat{z}$ and $\hat{r}$. The size versus time data is then fit with $\sigma(t)^2 = \sigma(0)^2 + t_m^2 (k_B T / m)$ to allow thermometry of the axial and radial directions.

\begin{figure}
	\includegraphics[width=\columnwidth]{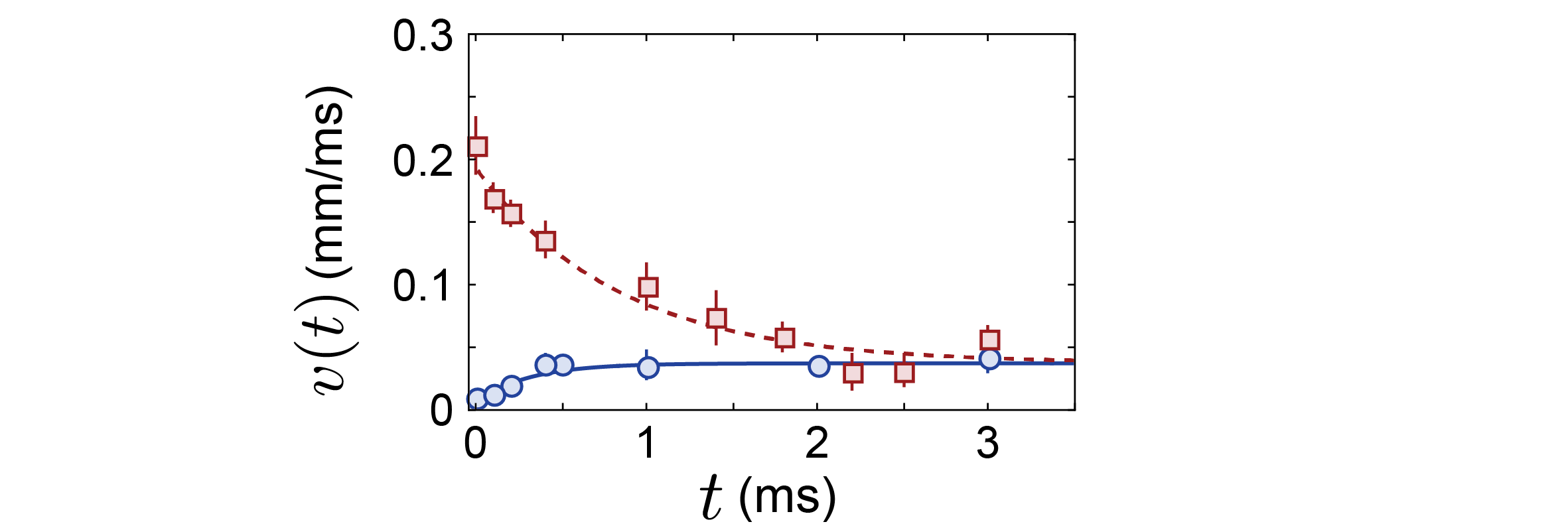}
	\caption{\label{fig:damping} Exemplary Velocity Damping Curves. The figure shows the center-of-mass velocity $v(t)$ versus the time $t$ that the $\Lambda$-BDM light is on. Blue circles (red squares) show data without (with) the push beam applied initially. The data without the push beam (blue circles) is used to determine $v_{\infty}$ (exponential fit shown by the blue solid line). The data with the push beam (red squares) is then fit with an exponential curve to determine $\alpha$ ($v_{\infty}$ fixed for the fit). The fitted values for $\alpha$ and $v_{\infty}$ are used to obtain $a(B)$ via $a(B) =  \alpha v_{\infty}$. The data shown was acquired at a magnetic field of $B = -1.14\,\text{G}$. For data shown in the main text, $v_{\infty}$ is determined from a measurement at a time $t$ well into the saturated regime ($t\gg 1/\alpha$).}
\end{figure}

\subsection{Measuring $F(B)$ for the $\Lambda$-BDM}
To measure the $\Lambda$-BDM force versus magnetic field profile $F(B) = m a(B)$, we first create a spatially small sample by loading molecules into a $\Lambda$-BDM at a magnetic gradient of $B'_z=22.9\,\text{G/cm}$. We then turn off the magnetic gradient and apply $\Lambda$-cooling in free space to obtain a low velocity spread. Subsequently, we apply a uniform magnetic field $\vec{B}=B\hat{x}$ and then optionally initialize the cloud with a finite velocity along $\hat{x}$ using a beam resonant with the $X-B$ transition (``push'' beam in Fig. \ref{fig: beam diagram}). We then switch on $\Lambda$-BDM light for a variable duration $t$. The resulting velocity of the cloud after $t$ is measured via time-of-flight with the following procedure. The cloud is allowed to evolve ballistically in the absence of light for a variable time $t_m$, after which it is imaged with resonant light for $1\,\text{ms}$. We perform a Gaussian fit to extract the center-of-mass position $x_{cm}$ versus time $t_m$. From a linear fit of $x_{cm}$ versus $t_m$, the velocity for each $\Lambda$-BDM on-time $t$ is obtained.

For each $B$, we assume that the equation of motion obeyed by the molecules is $m \ddot{x} = - m \alpha(B) \dot{x} +  F(B)$, consisting of a velocity damping term and a constant force.
Note that we assume that both $F$ and $\alpha$ are functions of $B$ only ($\alpha$ is assumed to be constant over the velocity range explored).
The equation-of-motion can be integrated to yield
\begin{equation}\label{eq: velocity decay}
v(t) = v_{\infty} + (v(0) - v_\infty) e^{-\alpha t},
\end{equation} 
where $v=\dot{x}$ and $v_\infty =  (F/m)/\alpha$ is the terminal velocity. By measuring $v$ as a function of the $\Lambda$-BDM on-time $t$ at different magnetic fields $B$, both $a(B)= F(B)/m$ and $\alpha(B)$ can be measured.

To obtain good estimates of $a(B)$ and $\alpha(B)$, we perform two sets of measurements with the molecules at two different initial velocities. First, after free space $\Lambda$-cooling, the cloud is already at rest ($v(0)=0$). Alternatively, by applying a push beam resonant with the $X-B$ transition for $4\,\mu\text{s}$, we can initialize the cloud at $v(0)\approx 200\,\text{mm/s}$. 

Because of the larger range of velocities explored, we find that precise estimates of $\alpha$ can be obtained from datasets with the push beam applied initially. On the other hand, precise estimates of $v_\infty$ are obtained from datasets where the molecules start from rest. We therefore use the push beam datasets to first obtain $\alpha$. Next, the terminal velocity $v_\infty$ of an initially stationary cloud (no push applied) is obtained using data where the $\Lambda$-BDM light is applied for much longer than $1/\alpha$. This provides a good estimate of $v_\infty$. Finally, the extracted value of $v_\infty$ is used to refit the push beam datasets to obtain a final estimate of the damping coefficient $\alpha$.

\section{BDM Properties}
In this section, we provide additional BDM properties.

\subsection{BDM Equilibration Time}
We find rapid equilibration times on the $5$ to $10$\,ms timescales for the four-frequency and two-frequency BDMs. At a gradient of $B_z' = \text{14.6}\,\text{G/cm}$, the four-frequency BDM reaches its equilibrium size over a $1/e$ time of $\sim10\,\text{ms}$ (Fig. \ref{fig:mot_timescale}(a)). For measurements of the equilibrium size and temperature of the four-frequency BDM, the molecules are held for $35\,\text{ms}$ at a gradient of $B_z' = 14.6\,\text{G/cm}$ to ensure that equilibrium is reached. For the two-frequency $\Lambda$-BDM ($B'_z=33.3\,\text{G/cm}$, $I$ optimized for maximum lifetime), an equilibrium size of $\sim 150\,\mu\text{m}$ is reached over $\sim5\,\text{ms}$ (Fig. \ref{fig:mot_timescale}(b)). We believe the quicker equilibration time in the high-gradient two-frequency configuration is due to a higher restoring force resulting from a higher value of $B_z'$ and $I$. We leave further exploration of equilibration times to future work.

We note that for the images shown in main text Fig.~1(b), the $\Lambda$-BDM is directly loaded from the free-space $\Lambda$-cooled cloud (without stages 1-3) and is therefore far from equilibrium initially. Notably, even at $12\,\text{ms}$ after initial loading (main text Fig.~1(b) lower right panel), which is longer than the measured $1/e$ equilibration time for small perturbations (Fig.~\ref{fig:mot_timescale}(b)), the cloud has still not fully reached its final equilibrium size. In particular, even though the axial width is close to its equilibrium value, the radial width is still a factor $\sim 2$ larger than the axial width; at equilibrium, the two widths are similar.

\subsection{$\Lambda$-BDM Densities and Phase-Space Densities versus Magnetic Gradient}
The phase-space density of a molecular cloud can be obtained from three quantities: molecule number, cloud size, and cloud temperature. We can measure the MOT sizes and temperatures along the axial and one radial direction. We assume the system is radially isotropic.

The number of molecules is measured by a $20\,\text{ms}$-long $\Lambda$-imaging pulse. After summing camera counts within an ROI, we convert the camera counts into the number of collected photons using known camera conversion factors. To convert the photon number into a molecule number, the scattering rate during imaging and the collection efficiency of our imaging system are needed. We calibrate the scattering rate by comparing the imaging lifetime with and without the $X{}^2\Sigma(v=3, N=1) \to A{}^2\Pi(v=2,J=1/2,+)$ repumper. Using the known vibrational branching ratios for the $A{}^2 \Pi \to X{}^2\Sigma$ decay pathway, we estimate a scattering rate of $6.7(16)\times10^4\,\text{s}^{-1}$. Combining this with the light collection efficiency of our imaging system, we obtain an estimate of the molecule number.

We measure the in-situ size of the BDM using a $2\,\text{ms}$ exposure. The size of the free-space $\Lambda$-cooled cloud is measured using a $\text{5}\,\text{ms}$ exposure. We extract cloud parameters using the fitting procedure described in Sec. \ref{sec:mot_size}. The temperature is measured via time-of-flight expansion (Sec. \ref{sec:temperature}), where the expanded cloud is imaged resonantly for $1\,\text{ms}$. 

\begin{figure}
	\includegraphics[width=\columnwidth]{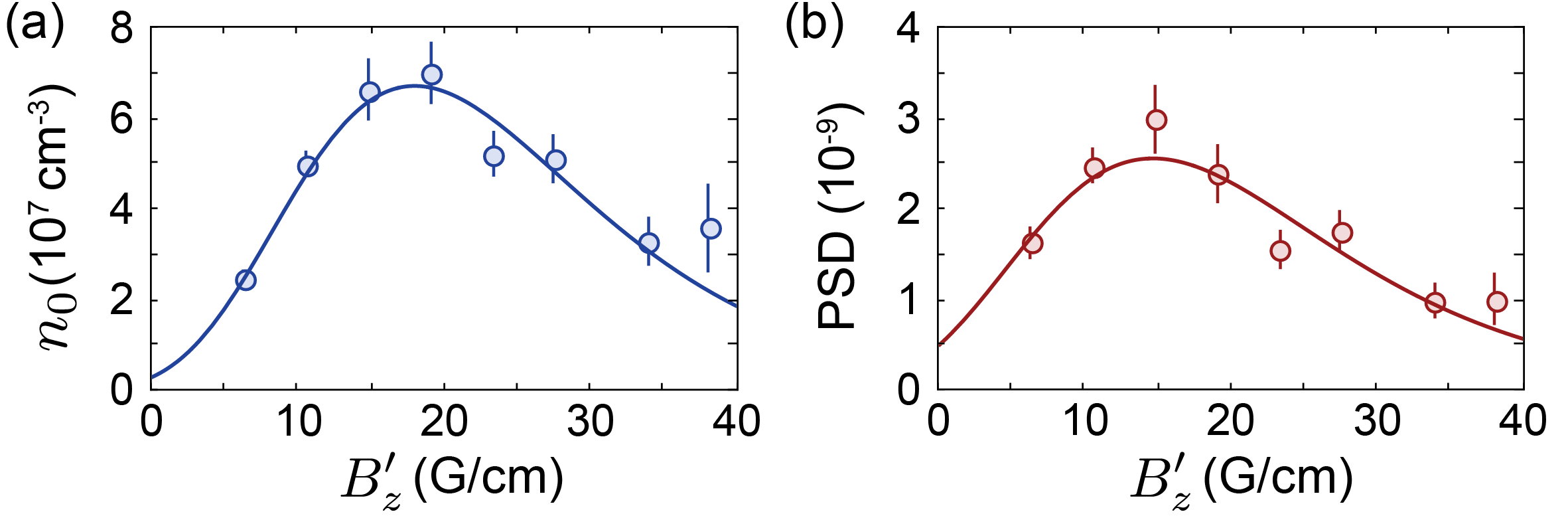}
	\caption{\label{fig:psd} $\Lambda$-BDM Density and PSD Dependences on Axial Magnetic Gradient $B_z'$. (a) Peak density $n_0$ versus $B_z'$. (b) Peak phase-space density $\text{PSD}$ versus $B_z'$. Solid lines are guides to the eye. }
\end{figure}

To obtain peak phase-space densities, we use the measured sizes, temperatures, and molecule number.  The phase-space density is given by $\text{PSD} = n_0 \lambda_{\text{dB}}^3$, where $n_0$ is the peak spatial density and $\lambda_{\text{dB}}$ is the thermal de Broglie wavelength given by $\lambda_{\text{dB}} = \sqrt{2\pi \hbar^2/(m k_B T)}$. The dependences of $n_0$ and $\text{PSD}$ of the $\Lambda$-BDM on the axial magnetic gradient $B_z'$ are shown in Fig.~\ref{fig:psd}. As mentioned in the main text, the highest peak density and the highest peak phase-space density are reached at different magnetic gradients.

\subsection{$\Lambda$-BDM properties at High Magnetic Gradients}
\begin{figure}
	\includegraphics[width=\columnwidth]{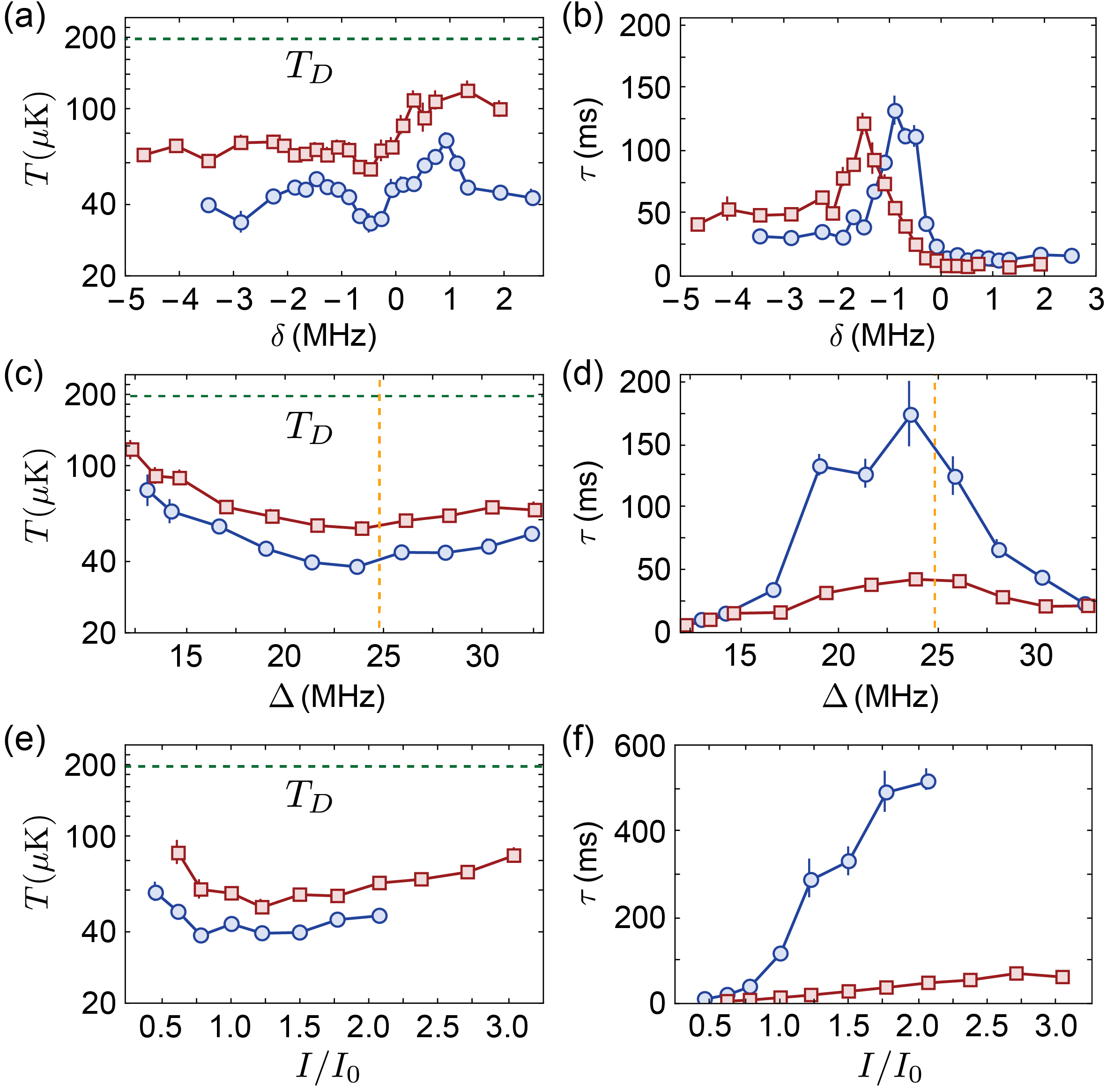}
	\caption{\label{fig:bdmparam} Mean Temperature $T$ and Lifetime $\tau$ Parameter Dependences in the $\Lambda$-BDM. (a,b) $T$ and $\tau$ versus two-photon detuning $\delta$. (c,d) $T$ and $\tau$ versus single-photon detuning $\Delta$. (e,f) $T$ and $\tau$ versus intensity $I/I_0$, where $I_0=5.8(2)\,\text{mW/cm}^2$. For all plots, blue circles show data for $B_z'=10.4\,\text{G/cm}$ (same as in main text Fig.~2), and red squares show data for $B_z'=33.6\,\text{G/cm}$. In (a,c,e), the horizontal dashed line indicates the Doppler temperature $T_D = 200\,\mu\text{K}$. In (c,d), the vertical dashed line shows the single-photon frequency detuning $\Delta$ at which the frequency component nominally addressing $F=2$ becomes resonant with $F=1^+$.}
\end{figure}

In Fig.~\ref{fig:bdmparam}, we show the dependence of the mean temperature $T$ and lifetime $\tau$ of the two-frequency BDM at moderate ($B_z'=10.4\,\text{G/cm}$, same as Fig.~2 in main text) and high gradient ($B_z'=33.6\,\text{G/cm}$). We find that the $\Lambda$ features, namely a dip in temperature and a peak in lifetime, are present even at high gradients. The two-photon detuning $\delta$ that produces the longest lifetime is shifted further from $\delta=0$ at the higher gradient configuration. We also find that with the higher gradient, the temperature is increased and the lifetime is decreased. Although the lifetime at the higher gradient can be extended by operating with higher beam intensities and at the shifted optimum in $\delta$, the lifetime remains $<100\,\text{ms}$ for all explored parameters.

\section{MOT Scaling Laws}
\subsection{Generalized Virial theorem}
\label{subsec:virial}
We first derive a generalized Virial theorem that allows us to equate the average kinetic energy $\langle \frac{1}{2} m v^2 \rangle $ with the average potential energy $\langle \frac{1}{2} k x^2 \rangle $ in steady-state for a damped harmonic oscillator that also undergoes diffusion in momentum space due to random momentum kicks from photon recoil.

Over timescales much longer than the spontaneous emission lifetime, the time-evolution of the system can be described by a Fokker-Planck-Kramers equation~\cite{Stenholm1986lasercooling,Parkins1992lasercooling,Devlin2018FPE}. We denote the probability of occupying a phase-space volume $d^3p\,d^3 x$ at time $t$ by $W(\vec{x},\vec{v},t)$. Assuming the diffusion constant is identical in all three directions and independent of velocity and position, the Fokker-Planck-Kramers equation reads
\begin{eqnarray}
& &\frac{\partial}{\partial t}  W(\vec{x},\vec{v},t) + \vec{v} \cdot \nabla_x W(\vec{x},\vec{v},t)  \nonumber \\
&=& - \nabla_v \cdot \left(\frac{\vec{F}(\vec{x},\vec{v})}{m} W(\vec{x},\vec{v},t)\right)+ \frac{D}{m^2} \nabla^2_v W(\vec{x},\vec{v},t),
\label{eq:FPE}
\end{eqnarray}
where $D= \hbar^2 k_L^2 \Gamma_{\text{sc}}/3$.

We next assume that the damping is identical in all directions, but that the spring constants are not necessarily identical. This gives $\vec{F}(\vec{x},\vec{v}) = -\sum_i k_i x_i \hat{x_i} - m \alpha \vec{v}$. We assume a Gaussian Ansatz for the steady-state distribution $W(\vec{x},\vec{v},t) =\prod_i \mathcal{N}_i \exp(- x_i^2/(2\sigma_i^2)) \exp (- mv_i^2/(2k_B T_i))$.
With this thermal Ansatz and in steady-state, the different directions decouple in Eq.~\ref{eq:FPE}. Considering one of the dimensions, we obtain
\begin{equation}
0- \frac{xv}{\sigma^2} = \left(-\frac{k}{m}x - \alpha v \right) \frac{m v}{k_B T}+ \alpha + \frac{D v^2}{(k_B T)^2}  - \frac{D}{m k_B T}
\end{equation}
Simplifying, we obtain
\begin{equation}
 \left(\frac{1}{\sigma^2}-\frac{k}{k_B T}\right) xv =  \left(\alpha - \frac{D }{m k_B T}\right) \frac{m v^2}{k_B T} -\left( \alpha - \frac{D}{m k_B T}\right).
\end{equation}
We therefore find that in steady-state
\begin{eqnarray}
k &=& \frac{k_B T}{\sigma^2}  \\
\frac{D}{m \alpha} &=& \frac{\hbar^2 k_L^2 \Gamma_{\text{sc}}}{3m \alpha} = k_B T
\end{eqnarray}

The first equation states that the ensemble averaged kinetic energy is equal to the ensemble averaged potential energy, which we will call the generalized Virial theorem. The second equation is an Einstein relation that relates momentum diffusion due to photon scattering with the temperature, which encodes the velocity spread.
The validity of both equations are verified experimentally, as described in Section~\ref{sec:crosschecks}. 

In general, for a power-law restoring force profile with linear friction ($\vec{F} = -\sum_ik_i |x_i|^n\hat{x}_i- m \alpha \vec{v}$), one can use an Ansatz of the form $W(\vec{x},\vec{v},t) =\prod_i \mathcal{N}_i \exp(- x_i^{n+1}/((n+1)\sigma_i^2)) \exp (- mv_i^2/(2k_B T_i))$ to solve the Fokker-Planck-Kramers equation. One then finds that $k= k_B T/\sigma^{n+1}$, and the ratio of the kinetic to potential energies coincides with the result of applying the Virial theorem to a system without damping. In particular, one finds that $\text{KE} = \frac{(n+1)}{2}\text{PE}$, where $\text{KE}$ and $\text{PE}$ denote the average kinetic and potential energies, respectively.

\subsection{MOT Size}
Using the generalized Virial theorem derived in the previous section, we find that the MOT size $\sigma$ satisfies
\begin{equation}
\sigma^2 = \expval{x^2} = k_B T / k.
\label{eq:springviral}
\end{equation}

This equation is used in two situations in the main text. First, this equation is used to infer the dependences of the MOT size and spring constant on the magnetic gradient. The restoring force in a MOT acquires its spatial dependence through the magnetic gradient, meaning that $F(x) = F(B(x))$. Near the center of the MOT, the system follows Hooke's law, $F= -k x$. Rewriting in terms of $F(B)$, we find that $F\approx \left.\frac{\partial F(B)}{\partial B} \right|_{B=0} B' x$. This implies that the spring constant scales linearly with the magnetic gradient, $k \propto B'$. 
Therefore, at constant temperature, the MOT size $\sigma$ scales inversely with the magnetic gradient as $\sigma \propto 1/\sqrt{B'}$.

Second, in the main text, we invert Eq.~\ref{eq:springviral} to extract an effective spring constant $k_{\text{eff}} = k_B T/\sigma^2$ along both the axial and radial directions using measured values of $T$ and $\sigma$ (shown in Fig.~3(d) of the main text). The deviation of $k_{\text{eff}}$ from being linear in $B_z'$ indicates deviations from Hooke's law.

\subsection{Mean $B$-Field}
The root-mean-square $B$-field experienced by the molecules is given by $B_{\text{rms}} = \sqrt{\expval{B^2}}$. At the center of a quadrupole field where the magnetic gradient is constant, we find that $B_{\text{rms}} \propto  B_z' \sigma$. Using the MOT size $\sigma$ obtained under the harmonic approximation described  in the previous subsection, we therefore find that $B_{\text{rms}} \propto  \sqrt{T} \sqrt{ B_z'}$. This shows that the molecules explore on average higher magnetic fields at higher MOT gradients.

\subsection{Effective Trap Potential}
To quantitatively analyze $\Lambda$-BDM trapping, we assume that the force $F(x)$ is derived from an effective conservative trapping potential $U(x)$ that is Gaussian. 
We further assume that $F$ is only a function of the local magnetic field $B$; hence the spatial dependence in the BDM arises from the magnetic gradient.

In other words, $F(x) = F(B(x)) = - \partial U(B(x))/\partial x = -U'(B) B'$, where $B'$ is the gradient. Assuming that $U(x)$ is Gaussian in $x$ and hence in $B$, the force can then be written as
\begin{equation}
F(B)  = - A B e^{- B^2/(2B_0^2)},
\end{equation}
where $B_0$ is the magnetic field over which the force is significant and $A$ is the overall amplitude. Using $B=B'x$ and integrating with respect to $x$ gives
\begin{equation}
U(x) = -\frac{A B_0^2}{B'} e^{- (B' x)^2/(2B_0^2)} = -U_0 e^{-x^2/(2 d^2)}
\end{equation}
which shows that the maximum trap depth is given by $U_0 = A B_0^2/B'$, with a trap size given by $d = B_0/B'$.

\section{Estimating BDM lifetime versus Magnetic Field Gradient}
In this section, we provide details on how we use the experimentally measured acceleration curve $a(B)$, damping curve $\alpha(B)$, and the measured dependences of $T$ and $\sigma$ on gradient $B_z'$ to estimate the lifetime of the $\Lambda$-BDM. In the main text, the resulting estimate is shown in the inset of Fig.~4(e).

As explained in the main text, we describe $a(B)$ via the force $F=m\,a(B)$ derived from an effective conservative Gaussian potential $U(x) = - U_0 \exp(-x^2/(2d^2))$, where $m$ is the molecular mass and $U_0$ is the peak effective trap depth. 

We assume that the spatial dependence arises from the magnetic gradient, which is expected to be accurate at low gradients where the effective motional magnetic fields are negligible compared to the mean magnetic field experienced by the molecules. We assume that the acceleration curves are identical for the radial and axial directions. This is partially justified by the similar values of the measured spring constants per gradient near $B_z'=0$ shown in Fig. 3(d) of the main text (See Section~\ref{sec:crosschecks}).

To estimate lifetimes, we first find a smooth interpolation of the measured values of the widths ($\sigma_r, \sigma_z$) and temperatures ($T_r$, $T_z$) versus the axial magnetic gradient $B_z'$. We find that a linear model for $T_r$ and $T_z$ describes the data well (Fig.~3(b) in main text). For the widths, we fit the observed widths to a phenomenological model of $\sigma(B_z') = a+b/x^c$. 

Next, we perform a Monte-Carlo simulation for a grid of axial gradient values up to $40\,\text{G}$. For each gradient, we create $10^4$ instances of axial positions $r$ and radial positions $z$. These radial (axial) positions satisfy a normal distribution with variance given by $\sigma_r$ $(\sigma_z)$ obtained from the smoothed model above. We treat the radial and axial positions independently. For each radial position, we compute the local magnetic field by multiplying by the radial magnetic gradient $B_r' = B_z'/2$. Similarly, for each axial position, we compute the local magnetic field by multiplying by the axial magnetic gradient $B_z'$. Having found the local magnetic field, we obtain the local damping rate $\alpha$ from $\alpha(B)$. We also determine the local trap depth $U_{0}$, which is obtained from the measured $a(B)$ curve. Note that $U_{0}$ depends on the local magnetic field and also the gradient. 

\begin{figure}[t]
	\includegraphics[width=\columnwidth]{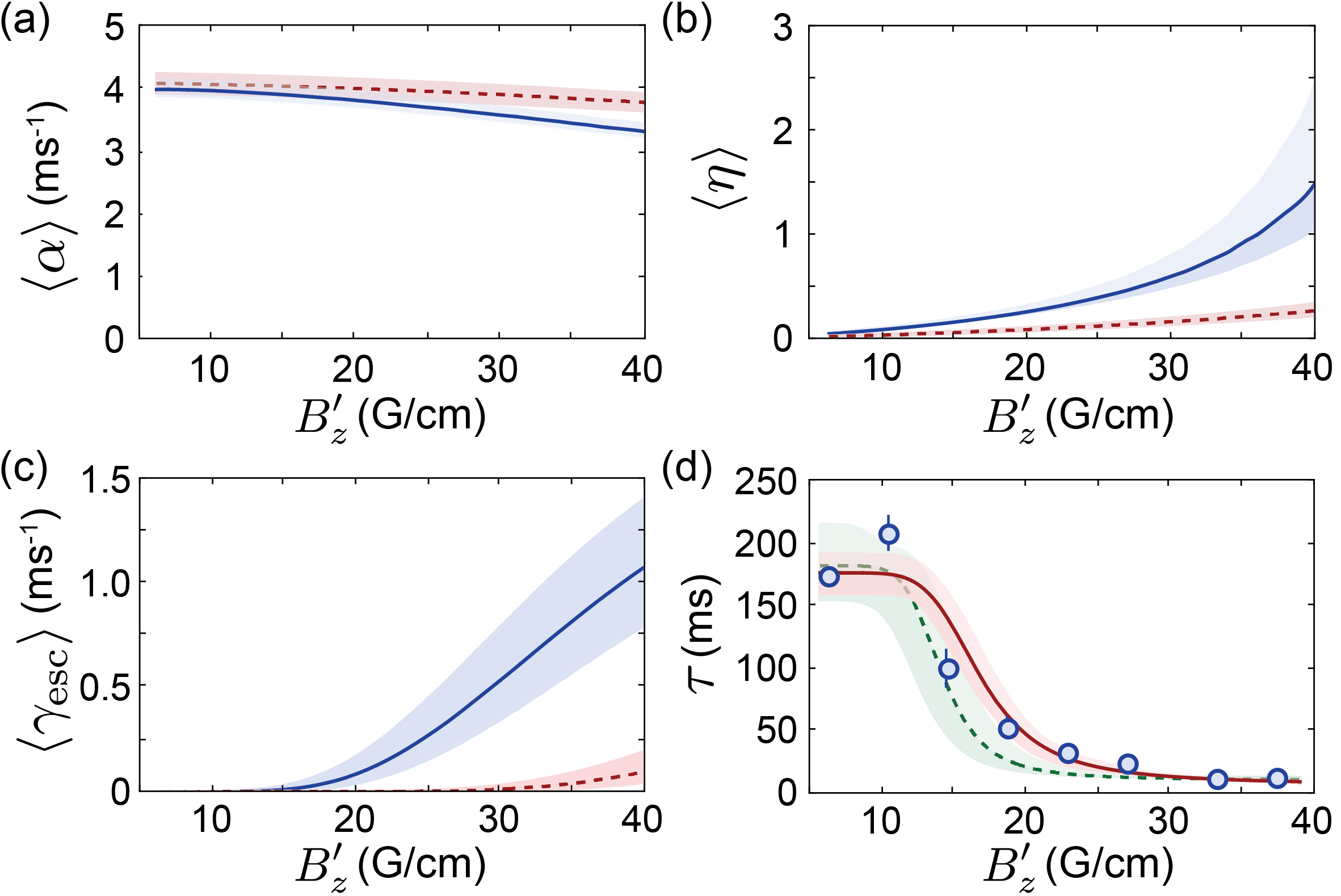}
	\caption{\label{fig:simresults} Lifetime Simulations Results. (a) Average damping $\langle \alpha \rangle$ for the axial (radial) direction shown by the blue solid line (red dashed line). (b) Average local temperature to trap depth ratio $\langle \eta \rangle$ for the axial (radial) direction shown by the blue solid line (red dashed line). (c) Escape rate $\langle \gamma_{\text{esc}} \rangle$ for the axial (radial) direction shown by the blue solid line (red dashed line). (d) Observed lifetime versus gradient $B_z'$ is shown by blue circles. The red solid line shows a fit to Monte-Carlo results with $C$ a fitting parameter; the green dashed line shows a fit to Monte-Carlo results with $C$ fixed to 1. For (a,b,c), the shaded regions represent the range covered due to uncertainty in the parameters used to model $T$ and $\sigma$ versus $B_z'$ (1 $\sigma$ in the model parameters). For (d), the shaded region includes additionally 1 $\sigma$ confidence interval for the fits. }
\end{figure}

Once $U_{\text{eff},r} (U_{\text{eff,z}})$ is computed for each radial (axial) position, we compute a list of $\eta_r = k_B T_r/U_{\text{eff},r}$ ($\eta_z= k_B T_z/U_{\text{eff},z}$) for each instance. For each value of axial gradient, $T_r$ an $T_z$ are obtained from the linear fits to measured data.
For each axial gradient value, which has $10^4$ radial and $10^4$ axial points, we can then perform ensemble averages for the mean damping rates $\langle \alpha \rangle$ and mean temperature-to-trap depth ratio $\langle \eta \rangle$. In Fig.~\ref{fig:simresults}(a)and Fig.~4(d) in the main text, we show the mean damping rates $\langle \alpha_r \rangle$ and $\langle \alpha_z \rangle$ along the radial and axial directions, respectively. In Fig.~\ref{fig:simresults}(b) and Fig~4(e) in the main text, we show the mean temperature-to-trap-depth ratio $\langle \eta_r \rangle$ and $\langle \eta_z \rangle$ along the radial and axial directions, respectively. Here angle brackets indicate an ensemble average over the different spatial positions within the trap. One can see that over axial gradients ranging from $5\,\text{G/cm}$ to $40\,\text{G/cm}$, $\langle \alpha_r \rangle$ and $\langle \alpha_z \rangle$ do not vary by more than $20\%$. On the other hand, $\langle \eta_r \rangle$ and $\langle \eta_z \rangle$ vary much more. $\langle \eta_z\rangle$ can even exceed unity above $30\,\text{G/cm}$.

Finally, to semi-quantitatively estimate the loss rate, we use the following picture. We assume that particles thermalize with the cooling light locally and that particles above the local trap depth are lost. For a thermal ensemble, the fraction of particles above the local trap depth is given by $\exp(-1/\eta)$. At the same time, the thermalization rate is given by the damping rate $\alpha$. Therefore, we can estimate the total escape rate as $\gamma = C\,\langle2\gamma_{\text{esc},r}+\gamma_{\text{esc},z}\rangle$, where $\gamma_{\text{esc},r}=\langle \alpha_r \exp(-1/\eta_r)\rangle$ is the radial escape rate, $\gamma_{\text{esc},z}=\langle \alpha_z \exp(-1/\eta_z)\rangle$ is the axial escape rate, and $C$ is a constant of order $\mathcal{O}(0.1)$. $C$ is used to phenomenologically capture the time needed to repopulate the tails of a thermal distribution, and to encode a slowing-down factor because hot particles have a finite probability of being recaptured before leaving the MOT. In Fig.~\ref{fig:simresults}(c), we show the mean escape rates $\langle \gamma_{\text{esc},z} \rangle$ and $\langle \gamma_{\text{esc},r} \rangle$ along the axial and radial directions, respectively. One observes that the axial escape rate is much higher and is the dominant escape channel. This is primarily because the axial gradient is twice as high as the radial gradient in a quadrupole magnetic field, leading to a lower axial trap depth. In addition, as shown in Fig.~3(b) in the main text, the axial temperature is also observed to be higher, partially contributing to the higher axial escape rate. 

In addition to the above loss mechanism, there is also a background loss due to off-resonant scattering into other vibrational and rotational states that are not repumped and also finite vacuum lifetime. This can be taken into account with a background loss rate $\gamma_{\text{bkg}}$. The total loss rate can then be estimated as
\begin{equation}
\gamma_{\text{tot}} = \gamma+ \gamma_{\text{bkg}}.
\end{equation}

At high loss rates, the measured lifetime is limited by the time it takes for a particle to escape the MOT region. To take this into account, we phenomenologically add to the estimated lifetime a constant diffusion time $\tau_{\text{diff}}$. The observed lifetime can then be estimated as 
\begin{equation}
\tau = \frac{1}{\gamma+ \gamma_{\text{bkg}}} + \tau_{\text{diff}}
\end{equation}

In Fig.~\ref{fig:simresults}(d), we show the observed lifetime $\tau$ as a function of the axial gradient $B_z'$, along with a fit with the model above (shown in red). We find qualitative agreement. The fit returns a background rate of $\gamma_{\text{bkg}}= 5.8(0.5)\,\text{s}^{-1}$, a diffusion time of $\tau_{\text{diff}}=5(2)\,\text{ms}$, and $C=0.21(8)$. We next check if these values are reasonable. Regarding $\gamma_{\text{bkg}}$, if we attribute this entirely to decay into un-repumped states during optical cycling, the corresponding number of photons scattered is $1.1(3) \times 10^5$. Regarding $\tau_{\text{diff}}$, the fitted value is not consistent with diffusive motion at $T\approx 100\,\mu\text{K}$. In detail, if one assumes diffusive motion, the time $t$ required to exit the BDM can be estimated as $t \sim d^2 \Gamma_{\text{sc}}/\bar{v}^2$, where $d$ is the size of the trap and $\bar{v}$ is the typical velocity. With $d=0.5\,\text{mm}$ and $T=100\,\mu \text{K}$, this timescale is $\sim 5\,\text{s}$. Possible explanations of the much shorter timescale could be because of an imbalance in beam intensities between opposing pairs of BDM beams that lead to a net force at large magnetic fields. With regards to $C$, as discussed previously, we expect $C$ to be on the order of 0.1. The fitted value is consistent with this expectation. In Fig.~\ref{fig:simresults}(d), we also show a fit of $\tau(B_z')$ to our model with $C$ fixed to 1 (shown in green), for which we still observe qualitative agreement.

\section{Estimating Collisional Loss}
In this section, we estimate the contribution of inelastic collisional loss to the total loss rate. The inelastic collisional loss rate for $X^2\Sigma\,(v=0,N=0,1)$ molecules at $\approx 40\,\mu \text{K}$ was reported to be near but below the universal value of $\beta = 3\times10^{-10}\,\text{cm}^3/\text{s}$~\cite{Cheuk2020collisions}. The light-assisted collisional loss rate for CaF molecules under $\Lambda$-cooling was reported to be $\sim 7$ times higher than the collisional loss rate of $X^2\Sigma\,(v=0, N=1)$ molecules~\cite{Anderegg2019Tweezer}. At our maximum observed density of $\approx 7\times 10^7\,\text{cm}^{-3}$, the estimated light-assisted collisional loss rate is $0.15\,\text{s}^{-1}$, which corresponds to a timescale of $\approx 7\,\text{s}$. This time-scale is much longer than any decay times observed for all explored BDM parameters in our work.

\section{Cross-checks}
\label{sec:crosschecks}
In this section, we describe two cross-checks of the acceleration curve $a(B_0)$ and damping curve $\alpha(B_0)$ measurements.

\subsection{Comparison of Spring Constant Obtained Globally and Directly via $a(B)$}
First, we check for consistency between the value of the spring constant obtained in Fig.~3(d) in the main text with the one obtained using the acceleration curve $a(B)$ in Fig.~4(b) in the main text.

In Fig.~3(d) of the main text, we interpret the data using a strongly damped harmonic oscillator model with random momentum kicks. As shown in Section~\ref{subsec:virial}, in steady-state, the average kinetic energy and the average potential energy are equal. The spring constant $k$ is related to the equilibrium size $\sigma$ and the temperature $T$ as described by Eq.~\ref{eq:springviral}. This expression can be rearranged to give
\begin{equation}
k_{\text{eff}} = \frac{k_B T}{\sigma^2},
\label{eq:kfig3}
\end{equation}
which is the expression used in the main text to extract an effective spring constant from measured values of $T$ and $\sigma$.  

Directly, the spring constant can also be obtained from the measured acceleration curve via 
\begin{equation}
k = - \frac{\partial F(B(x))}{\partial x} =  -\frac{\partial F}{\partial B} B',
\label{eq:kfig4}
\end{equation}
where $B'$ is the gradient. Near $x=0$, where $B \approx 0$ and Hooke's law holds, the spring constant obtained this way should coincide with the spring constant obtained using Eq.~\ref{eq:kfig3}.

Explicitly, because the axial gradient is twice as large as the radial gradient in a quadrupole magnetic field, we obtain
\begin{eqnarray}
k_z &=&-  \frac{\partial F(B)}{\partial B} B_z' \nonumber \\
k_r &=& -  \frac{\partial F(B)}{\partial B}\frac{B_z'}{2} \nonumber
\end{eqnarray}

We note that a key prediction of Eq.~\ref{eq:kfig4} is that $k$ is linear in the magnetic gradient at small values of $B$, where $\frac{\partial F}{\partial B}$ approaches a constant. As shown in Fig.~3(d) of the main text, indeed, $k= \frac{k_B T}{\sigma^2}$ is found to be linear in $B_z'$ at low gradients. 

As a cross-check, we compare the two approaches used to obtain $k$ at low gradients, Eq.~\ref{eq:kfig3}, which uses the generalized Virial theorem, and Eq.~\ref{eq:kfig4}, which makes use of direct measurement of $F(B) = m a(B)$. 

For quantitative comparison, we define $\kappa = \left.\frac{1}{m}\frac{\partial k}{ \partial B'}\right|_{B'=0}$. This expression can be used on data in Fig.~3d to extract $\kappa$. $\kappa$ can also be obtained from the measured force profile $a(B)$ using Eq.~\ref{eq:kfig4}, which gives $\kappa = -\left.\frac{\partial a(B)}{ \partial B}\right|_{B=0}$. From data in Fig.~3d, we obtain $\kappa = 0.31(11)\,(\text{mm/ms}^2)/\text{G}$ axially, and $\kappa = 0.25(6)\,(\text{mm/ms}^2)/\text{G}$ radially. From data of $a(B)$ shown in Fig.~4(b) of the main text, we obtain $\kappa =   0.165(29)\,(\text{mm/ms}^2)/\text{G}$. These values are consistent with each other ($1.4$ standard deviation difference).

Lastly, we compare our measured values for CaF with the values reported for a YO $\Lambda$-BDM at $B_z'=4\,\text{G/cm}$~\cite{Burau2023BDM}. At parameters that minimize temperature, work in YO reported axial and radial spring constants of $\approx 5 \times 10^{-21} \, \text{Nm}^{-1}$ and $\approx 2.5 \times 10^{-21} \, \text{Nm}^{-1}$, respectively. Using our measured acceleration curve $a(B)$ in CaF, we infer radial and axial spring constants at $B_z'=4\,\text{G/cm}$ of  $6.5(12) \times10^{-21}\, \text{Nm}^{-1}$ and $3.2(6) \times 10^{-21}\, \text{Nm}^{-1}$, respectively. These are comparable to the values obtained in YO. In terms of $\kappa$, the measurements in YO give $\kappa \approx 0.07\,(\text{mm/ms}^2)/\text{G}$ (both axially and radially), a factor of $\sim 3$ smaller than what we measured in CaF. The similar values for $k$ but differing values for $\kappa$ can be attributed to the larger mass of YO ($105\,\text{amu}$) compared to CaF ($ 59\,\text{amu}$).

\subsection{Validating the Einstein Relation Between Damping Rate and Scattering Rate}
As derived in Section IVA and used in previous work with BDMs of YO~\cite{Burau2023BDM}, one can use an Einstein relation to relate the momentum diffusion rate to the damping coefficient $\alpha$. Since the momentum diffusion rate is determined by the scattering rate $\Gamma_{\text{sc}}$, this produces an Einstein relation between $\Gamma_{\text{sc}}$ and $\alpha$, which reads
\begin{equation}
\frac{\hbar^2 k_L^2 \Gamma_{\text{sc}}}{3 m \alpha} = k_B T,
\vspace{0.0in}
\end{equation}
where $k_L$ is the wavenumber of the cooling light.

We can verify that the Einstein relation is satisfied by computing the ratio $\zeta = \frac{\hbar^2 k^2 \Gamma_{\text{sc}}}{3 m \alpha k_B T}$ from measured quantities. The relation is satisfied when $\zeta=1$. All the needed quantities on the right-hand side of the equation are experimentally measured at $B_z'=10.4\,\text{G/cm}$ in the $\Lambda$-BDM configuration ($\Delta = 23.8\,\text{MHz}$, $\delta = -0.7\,\text{MHz}$ and $I = 5.8(2)\,\text{mW/cm}^2$). In this configuration, we find $\zeta = 1.1(3)$, confirming that the Einstein relation is indeed satisfied within the measurement precision.

\bibliographystyle{apsrev4-1}